\documentclass[prd,eqsecnum,twocolumn,amsfonts,amssymb]{revtex4}

\usepackage{graphicx}

\usepackage{bm}

\newcommand{\beq}{\begin{equation}}
\newcommand{\eeq}{\end{equation}}
\newcommand{\beqs}{\begin{eqnarray}}
\newcommand{\eeqs}{\end{eqnarray}}
\newcommand{\lsim}{\mathrel{\raisebox{-
.6ex}{$\stackrel{\textstyle<}{\sim}$}}}

\begin{document}

\title{Constraints on Sterile Neutrinos in the MeV to GeV Mass Range} 

\author{D. A. Bryman$^{a,b}$ and R. Shrock$^c$}

\affiliation{(a) \ Department of Physics and Astronomy, University of British
Columbia, Vancouver British Columbia, V6T 1Z1, Canada}

\affiliation{(b) \ TRIUMF, 4004 Wesbrook Mall, Vancouver, British Columbia V6T
  2A3, Canada}

\affiliation{(c) \ C. N. Yang Institute for Theoretical Physics and
Department of Physics and Astronomy, \\
Stony Brook University, Stony Brook, New York 11794, USA }

\begin{abstract}

  A detailed discussion is given of the analysis of recent data to obtain
  improved upper bounds on the couplings $|U_{e4}|^2$ and $|U_{\mu 4}|^2$ for a
  mainly sterile neutrino mass eigenstate $\nu_4$. Using the excellent
  agreement among ${\cal F}t$ values for superallowed nuclear beta decay, an
  improved upper limit is derived for emission of a $\nu_4$.  The agreement of
  the ratios of branching ratios 
  $R^{(\pi)}_{e/\mu}=BR(\pi^+ \to e^+ \nu_e)/BR(\pi^+ \to \mu^+ \nu_\mu)$, 
  $R^{(K)}_{e/\mu}$, $R^{(D_s)}_{e/\tau}$, $R^{(D_s)}_{\mu/\tau}$, and 
  $R^{(D)}_{e/\tau}$, and the branching ratios $BR(B^+\rightarrow e^+\nu_e)$ 
  and $BR(B^+\rightarrow \mu^+\nu_\mu)$ decays with predictions of the 
  Standard Model, is utilized to derive new
  constraints on $\nu_4$ emission covering the $\nu_4$ mass range from MeV to
  GeV. We also discuss constraints from peak search experiments probing for
  emission of a $\nu_4$ via lepton mixing, as well as constraints from pion
  beta decay, CKM unitarity, $\mu$ decay, leptonic $\tau$ decay, and
  other experimental inputs.

\end{abstract}

\maketitle

% =======================================================================

\section{Introduction}
\label{intro_section}

In a recent paper \cite{ul}, we presented improved upper bounds on
the coupling $|U_{e4}|^2$ of an electron to a sterile neutrino $\nu_4$ from
analyses of data on nuclear and particle decays, for $\nu_4$ masses in the MeV
to GeV range, and pointed out new experiments that could
improve these constraints.  Here we give the details of our analysis that
yielded these constraints and also present a number of additional bounds on
sterile neutrino mixings, in particular, on the coupling $|U_{\mu 4}|^2$.

Neutrino oscillations and hence neutrino masses and lepton mixing have been
established and are of great importance as physics beyond the original Standard
Model (SM) \cite{davis}-\cite{accel}.  Most of the data from experiments with
solar, atmospheric, accelerator, and reactor (anti)neutrinos can be explained
within the minimal framework of three neutrino mass eigenstates with values of
$\Delta m^2_{ij} = m_{\nu_i}^2 - m_{\nu_j}^2$ given approximately by $\Delta
m^2_{21} = 0.74 \times 10^{-4}$ eV$^2$ and $|\Delta m^2_{32}| = 2.5 \times
10^{-3}$ eV$^2$, with normal mass ordering $m_{\nu_3} > m_{\nu_2}$ favored;
furthermore, the lepton mixing angles $\theta_{23}$, $\theta_{12}$, and
$\theta_{13}$ have been measured, with a tentative indication of a nonzero
value of the CP-violating quantity $\sin(\delta_{CP})$ (for compilations and
fits, see \cite{nu2018}-\cite{kopp_schwetz2018}).

The possible existence of light sterile neutrinos, in addition to the three
known neutrino mass eigenstates, is a fundamental question in particle physics.
These would have to be primarily electroweak-singlets (sterile), since the
invisible width of the $Z$ boson is consistent with being due to decays to
$\bar\nu_\ell \nu_\ell$, where $\nu_\ell=\nu_e, \ \nu_\mu$, and $\nu_\tau$,
corresponding to the known three SM fermion families\cite{zwidth}.  In the
presence of sterile neutrinos, the neutrino interaction eigenstates $\nu_e$,
$\nu_\mu$, and $\nu_\tau$ are linear combinations that include these additional
mass eigenstates.  In a basis in which the charged-leptons are simultaneously
flavor and mass eigenstates, the charged weak current has the form $J_\lambda =
\bar \ell \gamma_\lambda \nu_\ell$, where $\ell=e, \ \mu, \tau$ and
\beq
\nu_\ell = \sum_{i=1}^{3+n_s} U_{\ell i} \, \nu_i  \ , 
\label{nuell}
\eeq
where $n_s$ denotes the number of additional mass eigenstates. The
near-sterility of the $\nu_i$ with $4 \le i \le n_s$ is reflected in
small upper bounds on the corresponding $|U_{\ell i}|$.  
We will use the term ``sterile neutrino'' both in its precise sense as 
an electroweak-singlet interaction eigenstate and in a commonly used
approximate sense as the corresponding, mainly sterile, mass eigenstate(s) 
in this neutrino interaction eigenstate.  For technical simplicity, 
we will assume one heavy neutrino, $n_s=1$, with $i=4$; it is
straightforward to generalize to $n_s \ge 2$. Since a $\nu_4$ in the mass range
of interest here decays on a time scale much shorter than the age of the
universe, it is not excluded by the cosmological upper limit on the sum of
stable neutrinos, $\sum_i m_{\nu_i} \lsim 0.12$ eV
\cite{planck2018}. 

Possible sterile neutrinos are subject to many constraints from neutrino
oscillation experiments using solar and atmospheric neutrinos, accelerator and
reactor (anti)neutrinos, and kinematic effects in particle and nuclear decays,
as well as cosmological constraints. Bounds from the non-observation of
neutrinoless double beta decay are satisfied by assuming that $\nu_4$ is a
Dirac, rather than Majorana, neutrino. Although Majorana
neutrino masses have often been regarded as more generic, many
ultraviolet extensions of the SM contain additional gauge symmetries that
forbid Majorana mass terms, so that in these models, neutrinos are Dirac
fermions \cite{dirac}.  Much attention has been focused on possible sterile
neutrinos with masses in the eV region because of results from the LSND
\cite{lsnd} and Miniboone \cite{miniboone2018} experiments and possible
anomalies in reactor antineutrino experiments (recent reviews and discussions
include \cite{giunti_revs2019,conrad_shaevitz2019,liao}).  In addition to
eV-scale sterile neutrinos, there has also been interest in possible keV-scale
sterile neutrinos as warm dark matter, and in even heavier sterile neutrinos
with masses extending to the GeV range, and cosmological constraints on these
have been discussed \cite{cosm}-\cite{cosm5}. These cosmological constraints
involve assumptions about properties of the early universe.  One valuable
aspect of laboratory bounds on heavy neutrinos is that they are free of such
assumptions about the early universe.

Since sterile neutrinos violate the conditions for the
diagonality of the weak neutral current \cite{leeshrock77,sv80}, $\nu_4$ has
invisible tree-level decays of the form $\nu_4 \to \nu_j \bar \nu_i \nu_i$
where $1 \le i, j \le 3$ with model-dependent branching
ratios. Because our bounds are purely kinematic, they are complementary to
bounds from searches for neutrino decays, which involve model-dependent
assumptions on branching ratios into visible versus invisible final states.

This paper is organized as follows.  In
Sect. \ref{nucgen_section} we derive upper bounds on $|U_{e4}|^2$ from
nuclear beta decay data.  Sect.  \ref{pib_section} discusses pion beta
decay. Sect. \ref{ckm_section} considers connections of nuclear decay data with
the unitarity of the Cabibbo-Kobayashi-Maskawa (CKM) quark mixing matrix. In
Sect. \ref{peak_search_section} we discuss peak search experiments. In
Sects. \ref{e_mu_universality_section} and \ref{heavy_meson_decay_section} we
derive  upper bounds on lepton mixing matrix coefficients from two-body
leptonic decays of $\pi^+$, $K^+$, $D^+$, $D_s$, and $B^+$
mesons. Sects. \ref{mu_decay_section} and \ref{tau_decay_section} are devoted
to constraints from $\mu$ decay and leptonic $\tau$ decays. In
Sect. \ref{other_constraints_section} we briefly discuss other constraints on
sterile neutrinos.  Sect. \ref{conclusion_section} contains our conclusions.

% =====================================================================

\section{Limit on Emission of Massive Neutrinos in Nuclear Beta Decay}
\label{nucgen_section}

The emission of a heavy neutrino $\nu_j$ via lepton mixing and the
associated nonzero $|U_{ej}|^2$, with a mass in the keV-MeV region can 
be searched for in several ways using nuclear beta decays. If the
$\nu_j$ mass is less than the energy release $Q$ in a given beta
decay, its emission produces a kink in the Kurie plot. Ref.
\cite{shrock80} suggested a search for such kinks and used a
retroactive data analysis to set upper bounds on this type of
emission  via lepton mixing of neutrinos with kinematically
non-negligible masses in nuclear beta decays.  In standard notation,
$(Z,A)$ denotes a nucleus with $Z$ protons and $A$ nucleons.  For a
nuclear beta decay $(Z,A) \to (Z+1,A) + e^- + \bar\nu_e$ or
$(Z,A) \to (Z-1,A)+e^+ + \nu_e$ into a set of
neutrino mass eigenstates $\nu_i \in \nu_e$ with negligibly small masses
relative to the energy release in the decay plus a mass eigenstate
$\nu_4$ in $\nu_e$ with non-negligible mass, the differential decay rate is
\begin{widetext}
\beq
\frac{dN}{dE} = C \bigg [ (1-|U_{e 4}|^2)p E (E_0-E)^2
 + |U_{e 4}|^2 p E (E_0-E)\Big [(E_0-E)^2-m_{\nu_4}^2 \Big ]^{1/2} \,
\theta(E_0-E-m_{\nu_4}) \bigg ] \ ,
\label{nucldecrate}
\eeq
\end{widetext}
where $p\equiv |{\mathbf p}|$ and $E$ denote the 3-momentum and (total) energy
of the outgoing $e^\pm$ in the parent nucleus rest frame, $E_0$ denotes its
maximum energy for the SM case, the Heaviside $\theta$ function is defined as
$\theta(x)=1$ for $x > 0$ and $\theta(x)=0$ for $x \le 0$, and $C = G_F^2
|V_{ud}|^2 F_F |{\cal M}|^2/(2\pi^3)$, where ${\cal M}$ denotes the nuclear
transition matrix element, $V$ is the Cabibbo-Kobayashi-Maskawa (CKM) quark
mixing matrix, and $F_F$ is the Fermi function, which takes account of the
Coulomb interactions of the outgoing $e^\pm$. In general, there is also
a shape correction factor, but this is not important for the superallowed
decays considered here.  It is understood that if the decay is to an excited
state of the daughter nucleus rather than to its ground state, then there is a
corresponding reduction in the maximal value of $E_0$ relative to its value for
the decay to the ground state.  The kink in the Kurie plot arises as $E$
reaches the endpoint for the decay yielding a $\nu_4$ and the second term in
Eq. (\ref{nucldecrate}) vanishes.

Early bounds on $|U_{e4}|^2$ were set from searches for kinks in Kurie plots in
\cite{shrock80} and analyses of particle decays
\cite{shrock81a}-\cite{shrock_vpi}. Subsequently, dedicated experiments were
conducted to search for kinks in the Kurie plots due to possible emission of a
massive neutrino via lepton mixing for a number of nuclear beta decays over a
wide range of neutrino masses from O(10) eV to the MeV range.  For example, a
search for kinks in the Kurie plot in ${}^{20}$F beta decay reported in
Ref. \cite{deutsch1990} yielded an upper bound on $|U_{e4}|^2$ decreasing from
$5.9 \times 10^{-3}$ for $m_{\nu_4}=0.4$ MeV to $1.8 \times 10^{-3}$ for
$m_{\nu_4}=2.8$ MeV. (These and other upper bounds
discussed in this paper are at the 90 \% confidence level unless otherwise
stated.)  Some recent reviews of searches for sterile neutrinos in various
mass ranges include \cite{giunti_revs2019,conrad_shaevitz2019}, 
and \cite{helo2011}-\cite{boyarsky2019}. 

A general effect of the emission of a heavy neutrino $\nu_4$ in a nuclear beta
decay is to reduce the rate in a manner dependent on its mass, due to phase
space suppression of the decay, and, if it is too massive to be emitted, to
reduce the rate of the given decay by the factor $(1-|U_{e4}|^2)$.  Hence, in
addition to examination of Kurie plots for possible kinks, a powerful method to
constrain heavy neutrino emission, via lepton mixing, in nuclear beta decays is
to analyze the overall rates.  The apparent ($app$) rate, 
assuming no emission of a heavy neutrino, can be succinctly expressed as
\beq
\frac{dN}{dE}{}\Big |_{app}  \propto G_{F,app}^2 |V_{ud,app}|^2 F_{app} \ , 
\label{dnde_app}
\eeq
where $F_{app}=pE(E_0-E)^2$ is the SM kinematic function assuming no heavy
neutrino emission. Since, in general, the heavy neutrino would also
be emitted in $\mu$ decay, the measurement of the $\mu$ lifetime performed
assuming the SM would yield an apparent ($app$) value of the Fermi constant,
denoted $G_{F,app}$, that would be smaller than the true value
\cite{shrock81a,shrock81b,shrock_vpi}, $G_F$, given at tree level by 
\beq
\frac{G_F}{\sqrt{2}} = \frac{g^2}{8m_W^2} = \frac{g^2+g'^{\ 2}}{8m_Z^2} \ ,
\label{gf}
\eeq
where $g$ and $g'$ are the weak SU(2) and U(1)$_Y$ gauge couplings, and $m_W$
and $m_Z$ are the masses of the $W$ and $Z$ bosons. The apparent kinematic
function $F_{app}$ is larger than the true kinematic function indicated in the
square brackets in Eq. (\ref{nucldecrate}), which depends on $m_{\nu_4}$ and
$|U_{e4}|^2$. Since $G_{F,app}$ would be smaller than the true value of $G_F$,
while $F_{app}$ would be larger than the true $F$, the apparent value,
$|V_{ud,app}|^2$, extracted from a particular nuclear beta decay in the context
of the SM could be larger or smaller than the true value.  To avoid this
complication, we compare ratios of rates of different nuclear beta decays. In
these ratios, the factor $G_{F,app}^2$ cancels, so one can gain information
about the kinematic factor and hence about $|U_{e4}|^2$ as a function of
$m_{\nu_4}$.

The integration of $dN/dE$ over $E$ gives the kinematic rate factor $f$. The
combination of this with the half-life for the nuclear beta decay, $t \equiv
t_{1/2}$, yields the product $ft$. Incorporation of nuclear and radiative
corrections yields the corrected $ft$ value for a given decay, denoted ${\cal
  F}t$.  Conventionally, analyses of the most precisely measured superallowed
$0^+ \to 0^+$ nuclear beta decays have been used for many years to infer a
value of the weak mixing matrix element $|V_{ud}|$ \cite{th73,ht75}.  (In our
discussion of these fits, we will follow conventional notation and denote the
CKM mixing matrix factor as $V_{ud}$, with the implicit understanding that in
our present context with possible emission of a heavy neutrino $\nu_4$, this is
really $V_{ud,app}$.) In turn, these values of $|V_{ud}|$ extracted from
superallowed nuclear beta decays were used in early Cabibbo fits, e.g., 
\cite{roos74}, which were subsequently extended to the full CKM matrix
\cite{cab,ckm,blp79}.  The analyses of nuclear beta decay data have continued
up to the present with significant recent progress in precision 
\cite{strikman86}-\cite{tamu_conf}.

A first step in these analyses has been to establish the mutual
consistency of the ${\cal F}t$ values for these superallowed $0^+ \to 0^+$
decays. The emission of a $\nu_4$ with a mass $m_{\nu_4}$ of a
few MeV would have a different effect on the kinematic functions and integrated
rates for nuclear beta decays with different $Q$ (energy release) values and
would therefore upset this mutual consistency. Therefore, from this mutual
agreement of ${\cal F}t$ values, an upper limit on $|U_{e4}|^2$ can be derived
for values of $m_{\nu_4}$ in the MeV range, such that a $\nu_4$ could be 
emitted in some of these superallowed decays.
${\cal F}t$ is conventionally written as \cite{hardy_towner1990,cms2004,ms2006,hardy_towner2005,hardy_towner2015,ramsey_musolf,tamu_conf} 
\beq
{\cal F}t = \frac{K}{2G_V^2(1+\Delta_R^V)} \ , 
\label{ftft}
\eeq
where 
$K=2 \pi^3\ln2/m_e^5 = 0.81202776(9) \times 10^{-6} \ {\rm GeV}^{-4}-{\rm
  sec}$, $G_V=G_F|V_{ud}|$ and the radiative correction factor
$\Delta_R^V$ is transition-independent.
Ref. \cite{hardy_towner2015} obtains the 
average $\overline{{\cal  F}t}=3072.27 \pm 0.72$ sec.

The excellent mutual agreement between the ${\cal F}t$ values obtained from a
set of the most precisely measured superallowed $0^+ \to 0^+$ nuclear beta
decays, which involve only the vector part of the charged weak current, in
comparison with the value of $G_F$ obtained from muon decay, 
allows one to extract, in a self-consistent manner, a value of $|V_{ud}|$.  In
the 1990 study \cite{hardy_towner1990}, this yielded the result
$|V_{ud}|=0.9740 \pm 0.001$. At present, using a set of the fourteen most
precisely measured superallowed $0^+ \to 0^+$ nuclear beta decays, Hardy and
Towner have obtained the considerably more precise value
\cite{hardy_towner2018,hardy_pc} (denoted HT)
\beq
{\rm HT}: \quad |V_{ud}| = 0.97420(21)  \ . 
\label{vud_ht}
\eeq
Another recent estimate, in agreement with these, is $|V_{ud}|=0.97425(13)$
\cite{marciano_pc} (see also \cite{czms_july}). Using a different method for
calculating $\Delta_R^V$, Seng {\it et al.} \cite{ramsey_musolf} (denoted
SGPRM) obtain the slightly lower value
\beq
{\rm SGPRM}: \quad |V_{ud}| = 0.97370(14) \ , 
\label{vud_musolf}
\eeq
with a smaller reported uncertainty than in Eq. (\ref{vud_ht}).  As noted in
\cite{ramsey_musolf}, this lower value of $|V_{ud}|$ leads to tension with
first-row CKM unitarity.  Although the central values of $|V_{ud}|$ in
Eqs. (\ref{vud_ht}) and (\ref{vud_musolf}) differ, the bounds on $|U_{e4}|$ 
obtained below depend primarily on the precision in the mutual agreement of the
${\cal F}t$ values. 
The fourteen parent nuclei in the set used in \cite{hardy_towner2018} are
\beqs
&& {}^{10}{\rm C}, \ {}^{14}{\rm O}, \ {}^{22}{\rm Mg}, \ {}^{26m}{\rm Al}, \ 
{}^{34}{\rm Cl}, \ {}^{34}{\rm Ar}, \ {}^{38m}{\rm K}, \ {}^{38}{\rm Ca}
\cr\cr
&&
{}^{42}{\rm Sc}, \ {}^{46}{\rm V}, \ {}^{50}{\rm Mn}, \ {}^{54}{\rm Co}, \ 
{}^{62}{\rm Ga}, \ {\rm and} \ {}^{74}{\rm Rb} 
\label{superallowed_set} 
\eeqs
where the superscript $m$ refers to a metastable excited state. 
The maximal $Q$ value in this set is $Q=9.4$ MeV (${}^{74}$Rb) 
\cite{hardy_towner2015,nndc}.

\begin{figure}[hbt]
%\begin{center}
\centering
\includegraphics[height=12cm,angle=0,scale=0.5]{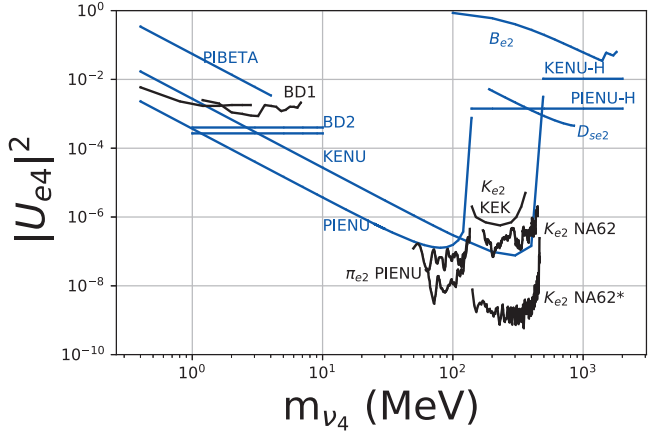}
%  \end{center}
\caption{90 \% C.L. upper limits on $|U_{e 4}|^2$ {\it vs.}
$m_{\nu_4}$ from various souces: PIBETA, pion beta decay (this work); 
BD1, previous limits from nuclear beta decay \cite{deutsch1990};
BD2, nuclear beta decay, based on our analysis using
\cite{hardy_towner2018} and \cite{ramsey_musolf}; 
PIENU and PIENU-H, the ratio
$\frac{BR(\pi^+ \to e^+\nu_e)}{BR(\pi^+ \to \mu^+ \nu_\mu)}$ in
the kinematically allowed and forbidden regions for $\nu_4$ emission
\cite{pienu2015}; 
$\pi_{e2}$ PIENU, $\pi^+ \to e^+ \nu_4$ peak searches 
(upper and lower curves from \cite{britton92} and \cite{pienu2018},
respectively); 
KENU and KENU-H, the ratio 
$\frac{BR(K^+ \to e^+\nu_e)}{BR(K^+ \to \mu^+ \nu_\mu)}$ in
the kinematically allowed and forbidden regions for $\nu_4$ emission; 
$K_{e2}$ KEK, $K^+ \to  e^+ \nu_4$ peak search \cite{nordkirchen};
$K_{e2}$ NA62, $K^+ \to  e^+ \nu_4$ peak search \cite{na62_2018}; and 
$K_{e2}$ NA62*, the preliminary upper limit from a $K^+ \to  e^+ \nu_4$ 
peak search \cite{na62_kaon2019}. Other bounds are denoted $D_{se2}$, from our
analysis of $\frac{BR(D_s^+ \to e^+\nu_e)}{BR(D_s^+ \to \tau^+\nu_\tau)}$, and 
$B_{e2}$, from our analysis of peak search data in $B^+ \to e^+ \nu_4$
\cite{park2016}. 
Our new bounds are colored blue (online), while previous bounds are colored 
black. See text for older bounds and further discussion.}
\label{Ue4_figure}
\end{figure}

The emission of a neutrino with a mass of order MeV in superallowed
nuclear beta decays would cause kinematic suppression depending on the
energy release $Q$ and the neutrino mass $m_{\nu_4}$, which would vary
from nucleus to nucleus owing to the different values of the phase
space factor in the third term, proportional to $|U_{e4}|^2$, in Eq.
(\ref{nucldecrate}).  Ref. \cite{strikman86} set upper limits on
$|U_{e4}|^2$ ranging from $3 \times 10^{-2}$ to $4 \times 10^{-3}$ for
$m_{\nu_4}$ from 0.5 MeV to 4.5 MeV, while
Ref. \cite{hardy_towner1990} obtained an upper bound on $|U_{e4}|^2$
ranging from $10^{-2}$ down to $2 \times 10^{-3}$ for $m_{\nu_4}$ from
0.5 to 2 MeV. Ref. \cite{deutsch1990} incorporated the phase space
integration for the massive-neutrino term proportional to $|U_{e4}|^2$
in Eq.  (\ref{nucldecrate}) for eight available superallowed beta
decays and then derived upper bounds on $|U_{e4}|^2$ from the
consistency of corrected ${\cal F}t$ values, depending
non-monotonically on $\nu_4$ masses from 1 to 7 MeV, with the results
$|U_{e4}|^2 < 1 \times10^{-3}$ to $|U_{e4}|^2 < 2 \times 10^{-3}$,
shown as BD1 in Fig. \ref{Ue4_figure}.

A measure of the mutual agreement among ${\cal F}t$ values of the superallowed
beta decays is the precision with which $|V_{ud}|^2$ is determined, so a
reduction in the fractional uncertainty of the value of $|V_{ud}|^2$ results in
an improved upper limit on $|U_{e4}|^2$. Let us denote this fractional
uncertainty from the $i$'th data analysis, as $[\delta^{(i)}
|V_{ud,i}|^2]/|V_{ud,i}|^2$. Then it follows that
\beq
\frac{\delta^{(2)} |U_{e4}|^2}{\delta^{(1)}|U_{e4}|^2} =
\frac{[\delta^{(2)} |V_{ud,2}|^2]/|V_{ud,2}|^2}
     {[\delta^{(1)} |V_{ud,1}|^2]/|V_{ud,1}|^2} \ . 
\label{deltaVud}
\eeq
The fractional uncertainties of $[\delta^{(2)} |V_{ud}|]/|V_{ud}| = 2 \times
10^{-4}$ and $1.4 \times 10^{-4}$ in Refs. 
\cite{hardy_towner2018,hardy_towner2015} and \cite{ramsey_musolf} are
improvements by the respective factors of 5 and 7.5 
relative to the inputs used in the 1990 studies
\cite{hardy_towner1990,deutsch1990}.

We use these improvements to infer respective improved upper bounds on
$|U_{e4}|^2$, following from the mutual agreement of the ${\cal F}t$ values
among the fourteen superallowed beta decays
\cite{hardy_towner2015,hardy_towner2018,ramsey_musolf}.  Using the HT value in
Eq. (\ref{vud_ht}), we find the upper bound
\beq
|U_{e4}|^2 \lsim 4 \times 10^{-4}
\label{ue4sq}
\eeq
for $\nu_4$ masses in the range from $m_{\nu_4} \simeq 1$ MeV to $m_{\nu_4}
\simeq 9.4$ MeV, as indicated in Fig.  \ref{Ue4_figure} (BD2, upper
line). Using the SGPRM value in Eq.  (\ref{vud_musolf}), we find
\beq
|U_{e4}|^2 \lsim 2.7 \times 10^{-4} \ , 
\label{ue4sq_rm}
\eeq
also shown in Fig. \ref {Ue4_figure} (BD2, lower line). Of course, the flat
line segments shown are approximations; the actual upper limits on $|U_{e4}|^2$
from the nuclear beta decay data are not precisely constant as a function of
$m_{\nu_4}$ over the range shown.  If the uncertainties in the ${\cal F}t$
values for each of the superallowed nuclear beta decays used for the overall
fit in \cite{hardy_towner2015,hardy_towner2018,ramsey_musolf} were equal, then
one could extend this analysis to derive an upper bound on $|U_{e4}|^2$ as a
function of $m_{\nu_4}$ in this range of 1 to 9.4 MeV.  However, this
condition, of equal precision for the measurement of the ${\cal F}t$ value of
each individual nuclear beta decay in this set, has not yet been achieved. For
this reason, we have conservatively presented our upper bounds (\ref{ue4sq})
and (\ref{ue4sq_rm}) as applying uniformly throughout the specified range $1 \
{\rm MeV} < m_{\nu_4} < 9.4$ MeV, i.e., as flat line segments in
Fig. \ref{Ue4_figure}.  

Since our bounds (\ref{ue4sq}) and (\ref{ue4sq_rm}) above do not
involve $|U_{\mu 4}|^2$,  they  complement the upper limits
on $|U_{e4}|^2$ derived from the measurement of the
ratio of decay rates $R^{(\pi)}_{e/\mu}=
\Gamma(\pi^+ \to e^+ \nu_e)/\Gamma(\pi^+ \to \mu^+ \nu_\mu)$ 
discussed in Sect. \ref{emuUGF} in the subset of the range of $\nu_4$ mass
values where they overlap, namely $1 \lsim m_{\nu_4} \lsim 10$ MeV. 

Other methods of determining $|V_{ud}|$ include pion beta decay (discussed in
Sec. \ref{pib_section}) and the neutron lifetime (which also has the
complication of involving the axial-vector part of the weak charged current),
but these are not as accurate as the determination from the superallowed $0^+
\to 0^+$ beta decays.

% ==========================================================================

\section{Limits from $\pi^+ \to \pi^0 e^+ \nu_e$ Decay}
\label{pib_section}

In this section we analyze limits on sterile neutrinos obtainable from pion
beta decay, $\pi^+ \to \pi^0 e^+ \nu_e$. The mass difference between the
charged and neutral pions is $\Delta_\pi = m_{\pi^+}-m_{\pi^0} = 4.5936 \pm
0.0005$ MeV \cite{pdg}.  It will be convenient to define
\beq
\epsilon_e = \frac{m_e^2}{\Delta_\pi^2} = 1.237 \times 10^{-2} \ .
\label{epsilon_e}
\eeq
If $\nu_e$ consists only of neutrino mass eigenstates with negligibly small
masses, then the Standard-Model expression for the decay rate, denoted
$\Gamma_{\pi\beta,SM}$, is \cite{kallen}
\beqs
\Gamma_{\pi \beta, SM} &=& \frac{G_F^2 |V_{ud}|^2 \Delta_\pi^5}{30\pi^3} \,
\bigg ( 1 - \frac{\Delta_\pi}{2m_{\pi^+}} \bigg )^3f(\epsilon_e)(1+\delta) \ ,
\cr\cr
&&
\label{gamma_pibeta}
\eeqs
where
\begin{widetext}
\beq
f(x) = (1-x)^{1/2} \Bigg [ 1 - \frac{9}{2}x - 4x^2 
 + \frac{15}{2} x^2 \ln \bigg ( \frac{1+\sqrt{1-x}}{\sqrt{x}} \bigg ) -
  \frac{3\Delta_\pi^2}{7(m_{\pi^+} + m_{\pi^0})^2} \Bigg ] 
\label{fpib}
\eeq
\end{widetext}
and $\delta$ incorporates radiative corrections,
calculated to be $\delta = 0.033$ \cite{sirlin,jaus}.
Note that the last term in the square brackets in Eq. (\ref{fpib})
is $-1.20 \times 10^{-4}$ and thus is much smaller than the
leading-$x$ terms. Neglecting this last term,   
the function $f(x)$ has the expansion
\beq
f(x) = 1 - 5x + \{ O(x^2), \ O(x^2\ln x) \} .
\label{fexp}
\eeq
If $\nu_e$ contains the known three neutrinos with masses that are
negligibly small for the kinematics here, together with an O(1) MeV
$\nu_4$, then the rate for pion beta decay has the form
\beq
\Gamma_{\pi\beta} =(1-|U_{e4}|^2) \Gamma_{\pi\beta,SM}
+ |U_{e4}|^2 \bar \Gamma_{\pi\beta,\nu_4}\theta(\Delta_\pi-m_e-m_{\nu_4}) \ ,
\label{gamma_pibeta_nu}
\eeq
where $\Gamma_{\pi\beta,\nu_4} \equiv|U_{e4}|^2 \bar
\Gamma_{\pi\beta,\nu_4}$ denotes the rate for the decay $\pi^+ \to
\pi^0 e^+ \nu_4$.  As in the case of nuclear beta decay, the
emission of the $\nu_4$ would produce a kink in the differential decay
distribution $d\Gamma_{\pi\beta}/dE_e$, where $E_e$ is the electron
energy.  In particular, while the maximum electron energy in the case
of emission of neutrinos with negligibly small masses is
\beq
E_{e,max,SM} = \frac{m_{\pi^+}^2 + m_e^2 - m_{\pi^0}^2}{2m_{\pi^+}} = 
4.01 \ {\rm MeV} \ , 
\label{emax_pib_sm}
\eeq
this is reduced to
\beq
E_{e,max,\nu_4} = \frac{m_{\pi^+}^2 + m_e^2 - 
(m_{\pi^0}+m_{\nu_4})^2}{2m_{\pi^+}} 
\label{emax_pib_nu4}
\eeq
in the $\pi^+ \to \pi^0 e^+ \nu_4$ decay.  However, in contrast to nuclear beta
decay, events ascribed to the decay $\pi^+ \to \pi^0 e^+ \nu_e$ were identified
by the diphoton decay of the $\pi^0$, and the $e^+$ energy was not
systematically measured, {\it e.g.}, in the PIBETA experiment at PSI
\cite{pocanic,pocanic_rev}. Hence, one could not do a kink search for this
decay, which would be quite difficult anyway because of the very small
branching ratio of $10^{-8}$ for pion beta decay.

However, one can use the comparison of the measured decay rate, or
equivalently, branching ratio for pion beta decay with the SM prediction to
obtain a limit on possible emission of a $\nu_4$.  We have
\beqs
\overline{BR}_{\pi\beta} &=& 
\frac{BR(\pi^+ \to \pi^0 e^+ \nu_e)}
     {BR(\pi^+ \to \pi^0 e^+ \nu_e)}_{SM} \cr\cr &
=& (1-|U_{e4}|^2) + |U_{e4}|^2 r_{\pi\beta,\nu_4} \ , 
\label{gamrat}
\eeqs
where $r_{\pi\beta,\nu_4}$ denotes the ratio of the kinematic
factor for the $\pi^+ \to \pi^0 e^+ \nu_4$ decay divided by that for
the decay into neutrinos of negligibly small mass, and, including radiative
corrections \cite{pocanic,pocanic_rev},
\beq
BR(\pi^+ \to \pi^0 e^+ \nu_e)_{SM} = (1.039 \pm 0.001) \times 10^{-8} \ .
\label{br_pibeta_sm}
\eeq
Defining $\epsilon_{\nu_4} = m_{\nu_4}^2/\Delta_\pi^2$, 
the function $r_{\pi\beta,\nu_4}$ can be approximated to leading order in 
$\epsilon_e$ and $\epsilon_{\nu_4}$ as 
\beq
r_{\pi\beta,\nu_4} \simeq 
\frac{1 - 5(\epsilon_e + \epsilon_{\nu_4})}{1-5\epsilon_e} \simeq
1-5\epsilon_{\nu_4} \ . 
\label{rho_approx}
\eeq
The current value listed by the Particle Data
Group, dominated by the PIBETA measurement \cite{pocanic,pocanic_rev}, is
\cite{pdg}
\beq
BR(\pi^+ \to \pi^0 e^+ \nu_e) = (1.036 \pm 0.006)\times 10^{-8} \ . 
\label{br_pibeta}
\eeq
This is in good agreement with the SM prediction (\ref{br_pibeta_sm}), yielding
\beq
\overline{BR}_{\pi/\beta} = 0.997 \pm 0.006 \ .
\label{pibeta_ratio_wrt_sm}
\eeq
From this we obtain the upper limit on $|U_{e4}|^2$ 
shown in Fig. \ref{Ue4_figure} as PIBETA.  As $m_{\nu_4}$
increases, and finally exceeds the value $m_{\pi^+}-m_{\pi^0}-m_e = 4.08$ MeV,
the decay $\pi^+ \to \pi^0 e^+ \nu_4$ is kinematically forbidden, and hence the
observed rate divided by the rate predicted in the SM with the usual mass
eigenstates in $\nu_e$ of negligibly small masses is reduced to the first term
in Eq. (\ref{gamrat}), namely $1-|U_{e4}|^2$.  The upper bounds on $|U_{e4}|^2$
from pion beta decay are less stringent than the bounds in Eqs. (\ref{ue4sq})
and (\ref{ue4sq_rm}).

% ==========================================================================

\section{Constraint from CKM Unitarity}
\label{ckm_section}

If the mass of $\nu_4$ were sufficiently large so that it could not be emitted
in any superallowed nuclear beta decays used in the determination of
$|V_{ud}|$, then, although there would still be mutual consistency in this
determination between the different superallowed nuclear decays, the result
would be a spurious apparent value of $|V_{ud}|^2$, namely $|V_{ud,app}|^2 =
|V_{ud}|^2 (1-|U_{e4}|^2)$ (where we again assume just one heavy neutrino).  In
turn, this would reduce the apparent value of $|V_{ud}|^2 + |V_{us}|^2 +
|V_{ub}|^2$ used to check the first-row unitarity of the CKM matrix. If one
uses the value of $|V_{ud}|$ in Eq. (\ref{vud_ht}), then the sum $|V_{ud}|^2 +
|V_{us}|^2 + |V_{ub}|^2$ is equal to unity to within the stated theoretical and
experimental uncertainties. Thus, this provides another constraint on
possible massive neutrino emission in the decays involved. Numerically, using
the value of $|V_{ud}|$ in Eq. (\ref{vud_ht}), together with the values
$|V_{us}|=0.2243(5)$ and $|V_{ub}|^2 = (1.55 \pm 0.28) \times 10^{-5}$ from
\cite{pdg}, Ref. \cite{hardy_towner2018} obtains
\beq
\Sigma \equiv |V_{ud}|^2 + |V_{us}|^2 + |V_{ub}|^2 = 0.99939(64) \ .
\label{ckmsumsq}
\eeq
The $|V_{ud}|^2$ term dominates both the sum and the uncertainty in
(\ref{ckmsumsq}). Thus, with the assumption of first-row CKM unitarity, this
also yields an upper limit on $|U_{e4}|^2$, depending on $m_{\nu_4}$ and
estimates of uncertainty in $|V_{us}|^2$.  If, on the other hand, one uses the
lower value of $|V_{ud}|$ in Eq. (\ref{vud_musolf}), then, as was observed in
\cite{ramsey_musolf}, there is tension with first-row CKM unitarity. However,
since the difference between the analyses in
\cite{hardy_towner2018,hardy_towner2015} and \cite{ramsey_musolf} is in the
value for the transition-independent correction term $\Delta_R^V$, this does
not upset the mutual agreement between the ${\cal F}t$ values, which was the
key input for the bound (\ref{ue4sq}).

% ============================================================================

\section{Constraints from Peak Search Experiments}
\label{peak_search_section}

It is also of considerable interest to discuss correlated limits on sterile
neutrinos from two-body leptonic decays of pseudoscalar mesons. Searches for
subdominant peaks in charged lepton momenta in two-body leptonic decays of
pseudoscalar mesons were suggested as a way to search for emission, via lepton
mixing, of a possible heavy neutrino $\nu_h$, and to set upper limits on the
associated couplings $|U_{\ell h}|^2$, also including effects on ratios of
branching ratios, in \cite{shrock80,shrock81a}.  These observations were
applied retroactively to existing data to derive such limits in
\cite{shrock80,shrock81a,shrock81b}.  In particular, the upper limit
$|U_{e4}|^2 \lsim 10^{-5}$ was obtained from retroactive analysis of data on
$K^+ \to e^+ \nu_e$ decays for $82 < m_{\nu_4} < 163$ MeV, and upper limits
on $|U_{\mu 4}|^2$ in the range $10^{-4} - 10^{-5}$ were obtained from data on
$\pi^+ \to \mu^+ \nu_\mu$( $\pi_{\mu 2}$) decay 
(Figs. 17, 22 in \cite{shrock81a}). An analogous
discussion of the emission of massive neutrino(s) in muon decay was given in
\cite{shrock81b,shrock82}, and an analysis of $\mu$ decay data was used in
\cite{shrock81b} to set upper limits on $|U_{e4}|^2$ and on $|U_{\mu4}|^2$ (see
Sect. \ref{mu_decay_section}).

Dedicated experiments have been carried out from 1981 to the present to search
for the emission, via lepton mixing, of a heavy neutrino in two-body leptonic
decays of $M^+ = \pi^+, \ K^+$ mesons and to search for effects of possible
heavy neutrinos on the ratio $BR(M^+ \to e^+\nu_e)/BR(M^+ \to \mu^+ \nu_\mu)$
\cite{abela81}-\cite{triumf_pimu2}. These have set very stringent bounds. Data
from the corresponding experiments with heavy-quark pseudoscalar 
mesons will be used below to derive new limits on sterile neutrinos. 
Some relevant properties of these experiments with two-body leptonic decays of
charged pseudoscalar mesons will be discussed next.  The peak search
experiments are quite sensitive to massive neutrino emission because one is
looking for a monochromatic signal and, furthermore, for a considerable range
of $m_{\nu_4}$ masses, there is a kinematic enhancement of the decays $M^+
\to e^+ \nu_4$ and $M^+ \to \mu^+ \nu_4$ relative to the
decays into neutrinos with negligibly small masses.

In the SM, the rate for the decay $M^+ \to \ell^+ \nu_\ell$ of a charged 
pseudoscalar $M^+$, where $M^+=\pi^+, \ K^+$, etc., and $\ell$ is a charged
lepton, is, to leading order, 
\beqs
\Gamma(M^+ \to \ell^+ \nu_\ell)_{SM} &=&
\frac{G_F^2 |V_{ij}|^2 f_M^2 m_M m_\ell^2}{8\pi} \,
 \bigg ( 1 - \frac{m_\ell^2}{m_M^2} \bigg )^2 \ , \cr\cr
&&
\label{gamma_sm}
\eeqs
where $V_{ij}$ is the relevant CKM mixing matrix element, $f_M$ is the
corresponding pseudoscalar decay constant (normalized such that $f_\pi = 130$
MeV), and we have used the fact that the three known neutrino mass eigenstates
$\nu_i$, $i=1,2,3$ in $\nu_\ell$ are negligibly small compared with $m_M$ for
all pseudoscalar mesons $M$.

However, because of lepton mixing, other decay modes may also
occur into some number of neutrinos with non-negligible masses. Focusing, as
above, on the case of a single heavy neutrino $\nu_4$, the SM rate is reduced
by the factor $(1-|U_{\ell 4}|^2)$ and there is another decay yielding the
heavy neutrino with rate,
\beq
\Gamma(M^+ \to \ell^+ \nu_4) = 
\frac{G_F^2 |V_{ij}|^2 |U_{\ell 4}|^2 f_M^2 m_M^3}{8\pi} \,
\rho(\delta_\ell^{(M)}, \delta_{\nu_4}^{(M)})  \ , 
\label{gamma_gen}
\eeq
in which the notation is as follows \cite{shrock80,shrock81a}: 
\beq
\delta^{(M)}_\ell =    \frac{m_\ell^2}{m_M^2} \ , \quad 
\delta^{(M)}_{\nu_4} = \frac{m_{\nu_4}^2}{m_M^2} \ , 
\label{deltas}
\eeq
\beqs
\rho(x,y) = f_{\cal M}(x,y) \, [\lambda(1,x,y)]^{1/2}  \ , 
\label{rhom}
\eeqs
where the factor $f_{\cal M}$ arises from the square of the matrix element
$\cal M$, 
and
\beq
f_{\cal M}(x,y) = x+y-(x-y)^2 \ .
\label{fm}
\eeq
In Eq. (\ref{rhom}), $\lambda(1,x,y)$ arises from the 
final-state two-body phase space, with 
\beq
\lambda(z,x,y) = x^2+y^2+z^2-2(xy+yz+zx) \ . 
\label{lambda1}
\eeq
Note that $\rho(x,y)$ has the symmetry property
\beq
\rho(x,y) = \rho(y,x) \ . 
\label{rhosym}
\eeq
In the SM case with zero or negligibly small neutrino masses, 
$\rho(x,0)=x(1-x)^2$. Here and below, it is implicitly
understood that $\rho(\delta^{(M)}_\ell,\delta^{(M)}_{\nu_4}) = 0$ if
$m_{\nu_4} \ge m_M-m_\ell$, since in this case the decay $M^+ \to \ell^+ \nu_4$
is kinematically forbidden.

There is a clear signature for the decay $M^+ \to \ell^+ \nu_4$ into a heavy
neutrino, namely the appearance of a monochromatic peak in the energy or
momentum distribution of the charged lepton below the dominant peak associated
with the emission of neutrino mass eigenstates of negligibly small mass.  The
energy and momentum of this additional peak, in the rest frame of the parent
meson $M$, are
\beq
E_\ell = \frac{m_M^2 + m_\ell^2 - m_{\nu_4}^2}{2m_M}
\label{eell}
\eeq
and 
\beq
p_\ell = |{\mathbf p}_\ell| = \frac{m_M}{2} \, 
\sqrt{\lambda(1,\delta_\ell^{(M)},\delta_{\nu_4}^{(M)})} \ . 
\label{pell}
\eeq
An experiment on the two-body leptonic decay of a
pseudoscalar meson $M^+ \to \ell^+ \nu_\ell$, searching for a subdominant peak
in the charged lepton momentum or energy distribution due to the decay $M^+ \to
\ell \nu_4$, is limited to the mass range $m_{\nu_4} < m_M - m_\ell$ for which
the decay is kinematically allowed.  It is also limited (i) to sufficiently
small $m_{\nu_4}$ such that the momentum or energy of the outgoing $\ell^+$ is
large enough so that the event will not be rejected by the lower cut used in
the event reconstruction and (ii) to sufficiently large $m_{\nu_4}$ so that the
subdominant peak can be resolved from the dominant peak.

The function $f_{\cal M}(\delta^{(M)}_\ell,\delta^{(M)}_{\nu_4})$ increases
from a minimum at $\delta_{\nu_4}=0$ to a maximum at
$\delta^{(M)}_{\nu_4}=(1/2)+\delta^{(M)}_\ell$, where it has the value
$2\delta^{(M)}_{\ell}+(1/4)$.  The maximum in $f_{\cal M}$ is in the physical
region if $m_\ell < (m_M/4)$. The ratio of the value of $f_{{\cal M},max}$
divided by $f_{\cal M}$ for emission of neutrinos of negligible mass is 
\beq
\frac{f_{{\cal M},max}}
     {f_{\cal M}(\delta^{(M)}_\ell,0)} = 
\frac{2\delta^{(M)}_\ell + \frac{1}{4} }
     {\delta^{(M)}_\ell (1-\delta^{(M)}_\ell) } \ . 
\label{fmratio}
\eeq
For decays in which $m_\ell << m_M$ and hence $\delta^{(M)}_\ell << 1$, this
produces a large enhancement, since 
\beqs
\frac{f_{{\cal M},max}}
     {f_{\cal M}(\delta^{(M)}_\ell,0)} &=&
\frac{1}{4\delta^{(M)}_\ell}\bigg [ 1 + O(\delta^{(M)}_\ell) \bigg ] \cr\cr
 &>>& 1 \quad.
\label{fmratio_approx}
\eeqs
For example, for $\pi_{e2}$, $K_{e2}$, $D_{e2}$, $(D_s)_{e2}$, and
$B_{e2}$ decays, this ratio (\ref{fmratio}) has the very large values $1.87
\times 10^4$, $2.33 \times 10^5$, $3.35 \times 10^6$, $3.71 \times 10^6$, and
$2.67 \times 10^7$, respectively.  Physically, these large enhancement factors
are due to the removal of the helicity suppression of the decay of the $M^+$
into a light $\ell^+$ and neutrinos $\nu_i$ with negligibly small masses.

It is convenient to define the ratio
\beq
\bar\rho(x,y) \equiv \frac{\rho(x,y)}{\rho(x,0)} = 
\frac{\rho(x,y)}{x(1-x)^2} \ . 
\label{rhobar}
\eeq
Thus, 
\beq
\frac{\Gamma(M^+ \to \ell^+ \nu_4)}
     {\Gamma(M^+ \to \ell^+ \nu_\ell)_{SM}} = 
\frac{|U_{\ell 4}|^2 \bar\rho(\delta_\ell^{(M)},\delta_{\nu_4}^{(M)})}
     {1-|U_{\ell 4}|^2} \ . 
\label{br_ratio}
\eeq
Note that the dominant radiative corrections divide out between the numerator
and denominator of Eq. (\ref{br_ratio}). 
Since a given value of lepton momentum $p_\ell$ is uniquely determined by
$m_{\nu_4}$ for a given pseudoscalar meson $M$, a null observation of an 
additional peak in an experiment and hence an upper limit on the ratio
$\Gamma(M^+ \to \ell^+ \nu_4)/\Gamma(M^+ \to \ell^+ \nu_\ell)_{SM}$ at a
particular $p_\ell$ yields an upper limit on $|U_{\ell 4}|^2$ for the
corresponding value of $m_{\nu_4}$. Solving Eq. (\ref{br_ratio}) for 
$|U_{\ell 4}|^2$ gives 
\beq
|U_{\ell 4}|^2 = \frac{\frac{\Gamma(M^+ \to \ell^+ \nu_4)}
     {\Gamma(M^+ \to \ell^+ \nu_\ell)_{SM}}}
{\frac{\Gamma(M^+ \to \ell^+ \nu_4)}
     {\Gamma(M^+ \to \ell^+ \nu_\ell)_{SM}} + 
\bar\rho(\delta_\ell^{(M)},\delta_{\nu_4}^{(M)})} \ . 
\label{usq_eq}
\eeq
Hence, denoting $\Gamma(M^+ \to \ell^+ \nu_4)_{ul}$ as the upper limit on 
$\Gamma(M^+ \to \ell^+ \nu_4)$, one has the resultant upper limit on 
$|U_{\ell 4}|^2$:
\beq
|U_{\ell 4}|^2 < \frac{\frac{\Gamma(M^+ \to \ell^+ \nu_4)_{ul}}
     {\Gamma(M^+ \to \ell^+ \nu_\ell)_{SM}}}
{\frac{\Gamma(M^+ \to \ell^+ \nu_4)}
     {\Gamma(M^+ \to \ell^+ \nu_\ell)_{SM}} + 
\bar\rho(\delta_\ell^{(M)},\delta_{\nu_4}^{(M)})} \ . 
\label{usq_upper_full}
\eeq
Provided that $|U_{\ell 4}|^2 << 1$, the right-hand side of
Eq. (\ref{br_ratio}) is, to a good approximation, equal to $|U_{\ell 4}|^2
\bar\rho(\delta_\ell^{(M)},\delta_{\nu_4}^{(M)})$, so that the upper limit
(\ref{usq_upper_full}) simplifies to
\beq
|U_{\ell 4}|^2 < \frac{\frac{\Gamma(M^+ \to \ell^+ \nu_4)_{ul}}
     {\Gamma(M^+ \to \ell^+ \nu_\ell)_{SM}}}
{\bar\rho(\delta_\ell^{(M)},\delta_{\nu_4}^{(M)})} \ . 
\label{usq_upper}
\eeq

The large values of $\bar\rho(\delta_e^{(M)},\delta_{\nu_4}^{(M)})$ decays over
a substantial part of the kinematically allowed range of $m_{\nu_4}$ mean that
$M^+ \to e^+ \nu_4$ decays are quite sensitive to the possible emission of a
heavy $\nu_4$. With fixed $x$, the function $\bar\rho(x,y)$ has the following
Taylor series expansion in $y$ for small $y$: 
\beq
\bar\rho(x,y) = 1 + \bigg [ \frac{1-3x^2}{x(1-x)^2} \bigg ] y + O(y^2) \ . 
\label{rhobar_taylor}
\eeq
The derivative of $\bar\rho(x,y)$ with respect to $y$ is 
\begin{widetext}
\beq
\frac{d\bar\rho(x,y)}{dy}=\frac{1}{x(1-x)^2} \, \frac{d\rho(x,y)}{dy} 
=\frac{1-x-5y-3x^2+7y^2-4xy+9xy(y-x)+3(x^3-y^3)}
{x(1-x)^2 [\lambda(1,x,y)]^{1/2}} \ . 
\label{drho_dy}
\eeq
\end{widetext}
Hence, 
\beq
\frac{d\bar\rho(x,y)}{dy}\bigg |_{y=0} =  \frac{1-3x^2}{x(1-x)^2} \ .
\label{drhobar_dy}
\eeq
In our application, 
\beq
x=\delta^{(M)}_\ell \  {\rm and} \quad y=\delta^{(M)}_{\nu_4} \ . 
\label{xy}
\eeq
For $M^+ \to \ell^+ \nu_4$ decays such that $\delta^{(M)}_\ell << 1$, which
include all of the $M^+ \to e^+ \nu_4$ decays, the derivative
(\ref{drhobar_dy}) is $[d\bar\rho(x,y)/dy]_{y=0} = x^{-1}[1+O(x)]$, i.e., 
\beq
\frac{d\bar\rho(\delta^{(M)}_\ell,\delta^{(M)}_{\nu_4})}{d\delta_{\nu_4}} 
{}\bigg |_{\delta^{(M)}_{\nu_4}=0} = \frac{1}{\delta^{(M)}_\ell} \bigg [ 1 + 
O(\delta^{(M)}_\ell) \bigg ] >> 1 \ . 
\label{drhobar_dy_me2}
\eeq
Hence, in $M^+ \to e^+ \nu_4$ decays, as $\delta^{(M)}_{\nu_4}$ increases from
 0, $\bar\rho(\delta^{(M)}_\ell,\delta^{(M)}_{\nu_4})$ increases very rapidly
   from unity to values $>> 1$. 

For a given $x$, the maximal value of $\bar\rho(x,y)$, as a function of $y$
occurs where $d\rho(x,y)/dy=0$, or equivalently, $d\bar\rho(x,y)/dy=0$ in
the physical region.  The value of $y$ at this maximum is given by the
solution for $y$, of the equation
\beqs
&& 3y^3-(9x+7)y^2 + (9x^2+4x+5)y\cr\cr
&& +(-3x^3+3x^2+x-1)=0 \ . 
\label{yeq}
\eeqs
In $M^+ \to e^+ \nu_4$ decays, $x = \delta^{(M)}_e << 1$, so that, to a
very good approximation, Eq. (\ref{yeq}) reduces to the equation
$(3y-1)(y-1)^2=0$. The relevant solution of this equation, giving the value of
$y$ at which $\rho(x,y)$ and $\bar\rho(x,y)$ reach their respective maxima if
$x << 1$, is
\beqs
y_{\bar\rho_{max}} &=& \frac{1}{3}, \quad i.e., \cr\cr
           m_{\nu_4} &=& \frac{m_M}{\sqrt{3}} \quad. 
\label{y_rhobar_max_xsmall}
\eeqs
Then (with $x << 1$), 
\beq
\bar\rho(x,1/3) = \frac{4}{27x} \, \bigg [ 1 + \frac{13}{2} x + O(x^2) \bigg ] 
\ , 
\label{rhobarmax_taylor}
\eeq
so
\beqs
[\bar\rho(x,y)]_{max} \simeq \bar\rho(x,1/3) = \frac{4}{27x} \ , 
\label{rhobarmax_xsmall}
\eeqs
which is $>> 1$. In Table \ref{rhobar_values} we list
the maximal values of $\bar\rho(x,y)=
\bar\rho(\delta_\ell^{(M)},\delta_{\nu_4}^{(M)})$ for the various pseudoscalar
mesons $M^+$ considered here, and for $\ell=e, \ \mu$, together with the
respective values of $m_{\nu_4}$ where these maxima occur.  Particularly large
maximal values of the $\bar\rho$ function occur for heavy-quark pseudoscalar
mesons, including $1.98 \times 10^6$, $2.20 \times 10^6$, and $1.58 \times
10^7$ for the $D^+ \to e^+ \nu_4$, $D_s^+ \to e^+ \nu_4$, and $B^+ \to e^+
\nu_4$ decays, respectively. As is evident from this table, these maximal
values are only slightly less than the maximal values of $\bar f_{\cal
  M}(\delta_\ell^{(M)},\delta_{\nu_4}^{(M)})$ mentioned above.  This is due to
the slow falloff of the two-body phase space factor $[\lambda(1,x,y)]^{1/2} = 
[\lambda(1,\delta_\ell^{(M)},\delta_{\nu_4}^{(M)})]^{1/2}$ with increasing
$m_{\nu_4}$.  To see this, let us define, as in \cite{shrock81a}, the 
ratio of the phase space factor divided by its value for zero neutrino mass, 
\beq
[\bar \lambda(1,x,y)]^{1/2} \equiv 
\frac{[\lambda(1,x,y)]^{1/2}}{[\lambda(1,x,0)]^{1/2}} = 
\frac{[\lambda(1,x,y)]^{1/2}}{1-x} \ . 
\label{lambdabar}
\eeq
This has the Taylor series expansion 
\beq
[\bar \lambda(1,x,y)]^{1/2} = 1 - \frac{(1+x)}{(1-x)^2} \, y + O(y^2)
\label{lam_taylor}
\eeq
for small $y$.  Hence, the phase space function normalized to its value for
zero neutrino mass, {\it i.e.}, 
$[\bar\lambda(1,\delta_\ell^{(M)},\delta_{\nu_4}^{(M)})]^{1/2}$, decreases 
from unity rather slowly for 
small $\delta_{\nu_4}^{(M)}$. The maximal value of $y=\delta_{\nu_4}^{(M)}$ is 
\beqs
y_{max}&=& (1-\sqrt{x})^2, \quad i.e., \cr\cr
(\delta^{(M)}_{\nu_4})_{max} &=& \Big (1-\sqrt{\delta^{(M)}_\ell} \ \Big )^2 .
\label{ymax}
\eeqs
For fixed 
$x=\delta^{(M)}_\ell$, as $y=\delta^{(M)}_{\nu_4}$ approaches $y_{max}$ from
below, the phase space factor $[\lambda(1,x,y)]^{1/2} \to 0$, and hence so do
$\rho(x,y)$ and $\bar\rho(x,y)$. From the factorized expression 
\beqs
\lambda(1,x,y) &=& (1+\sqrt{x}+\sqrt{y} \ )(1+\sqrt{x}-\sqrt{y} \ )  
\cr\cr
&\times& (1-\sqrt{x}+\sqrt{y} \ )(1-\sqrt{x}-\sqrt{y} \ ) \cr\cr
&&
\label{lamfac}
\eeqs
it follows that as $y \to y_{max}$ from below, $[\lambda(1,x,y)]^{1/2}$
vanishes like $2 x^{1/4} \, (y_{max}-y)^{1/2}$. Hence, 
\beq
\frac{d\bar\rho(x,y)}{dy} \to 
-\frac{2x^{3/4}}{\Big [1-\frac{y}{y_{max}} \Big ]^{1/2} } 
\quad {\rm as} \ \ y \to y_{max} \ .
\label{drho_dy_ymax}
\eeq
From Eq. (\ref{drho_dy_ymax}), it follows that for
any physical value of $x$, as $y \to y_{max}$ from below, $\rho(x,y)$ and
$\bar\rho(x,y)$ approach 0 with a negatively infinite slope. 

For fixed $x$, over almost all of the kinematically allowed region in $y$,
the reduced function $\bar\rho(x,y)$ is larger than 1. The fact that
$\bar\rho(\delta_e^{(M)},\delta_{\nu_4}^{(M)}) > 1$ up to values of $m_{\nu}$
extremely close to its upper endpoint is understandable in view of the property
embodied in Eq. (\ref{drho_dy_ymax}),
that this function approaches zero with a slope that approaches $-\infty$,
i.e., nearly vertically, as $\delta^{(M)}_{\nu_4} \to 
(\delta^{(M)}_{\nu_4})_{max}$. For example,
in the $M^+ \to e^+ \nu_4$ decay, with $M^+=\pi^+$ or $K^+$,
$\bar\rho(\delta_e^{(M)},\delta_{\nu_4}^{(M)}) > 1$ for all $m_{\nu_4} > 0$ up
to values that are within 0.015 MeV of the respective kinematic endpoints
$m_{\pi^+}-m_e=139.059$ MeV and $m_{K^+}-m_e= 493.156$ MeV.  At these
respective values of $m_{\nu_4}$, the momentum of the $e^+$ is very small,
namely 0.125 MeV, which would be below the lower cutoff for such an event to be
accepted as a $\pi_{e2}$ or $K_{e2}$ event. Similar comments apply for the
leptonic decays of heavy-quark pseudoscalar mesons. We will use this property
in the limits that we derive below on $|U_{e4}|^2$.

\begin{figure}[hbt]
% \begin{center}
\centering
\includegraphics[height=12cm,angle=0,scale=0.5]{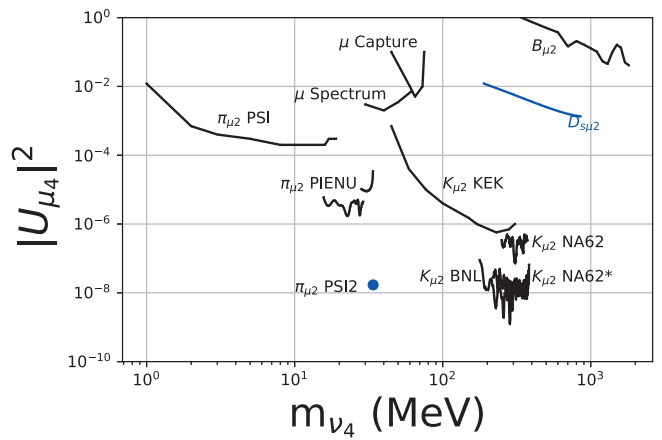}
%  \end{center}
\caption{Best 90 \% C.L. upper limits on $|U_{\mu 4}|^2$ {\it vs.}
$m_{\nu_4}$ from various experiments: 
$\pi^+\rightarrow \mu^+ \nu_4$ peak searches, labeled as follows: 
$\pi_{\mu 2}$ PSI \cite{daum87}, $\pi_{\mu 2}$ PSI2 \cite{daum2000}, 
$\pi_{\mu 2}$ PIENU \cite{triumf_pimu2}; 
$K^+\rightarrow \mu^+ \nu_4$ peak searches: 
$K_{\mu 2}$ KEK \cite{hayano82,nordkirchen},
$K_{\mu 2}$ BNL \cite{bnl949}, 
$K_{\mu 2}$ NA62 \cite{na62_2018}, and the preliminary limit 
$K_{\mu 2}$ NA62* \cite{na62_kaon2019}. Other limits include 
$\mu$ spectrum \cite{shrock81b};  
$\mu$ capture \cite{deutsch83}; 
a $B^+\rightarrow \mu^+ \nu_4$ peak search denoted $B_{\mu 2}$ 
\cite{park2016}; and our analysis of 
$\frac{BR(D_s^+ \rightarrow \mu^+ \nu_\mu)}
      {BR(D_s^+ \rightarrow \tau^+ \tau_\mu)}$, labeled $D_{s \mu 2}$,
Our new bounds are colored blue (online) while previous bounds are colored 
black. See text for previous bounds and futher discussion.}
\label{Umu4_figure}
\end{figure}

Recent bounds from $\pi_{\ell 2}$ and $K_{\ell 2}$ peak search experiments
include those from the searches for $\pi^+ \to e^+ \nu_h$ and 
$\pi^+ \to \mu^+ \nu_h$ decays by the PIENU
experiment at TRIUMF \cite{pienu2015,pienu2018,triumf_pimu2}, for 
$K^+ \to \mu^+ \nu_h$ decay in the E949 experiment at BNL \cite{bnl949}, and 
for the $K^+ \to \mu^+ \nu_h$ and $K^+ \to e^+ \nu_h$ decays 
in the NA62 experiment at CERN \cite{na62_2017,na62_2018,na62_kaon2019}, 
where $\nu_h \equiv \nu_4$ in our notation.  From
the various $\pi_{e2}$, $\pi_{\mu 2}$, $K_{e 2}$, and $K_{\mu 2}$ 
peak search experiments, some upper bounds include 
\begin{itemize}

\item 
$|U_{e4}|^2 \lsim 10^{-7} - 10^{-8}$ for $50 \ {\rm MeV} \ < m_{\nu_4} < 135$ 
MeV \cite{pienu2015,pienu2018};  

\item 
$|U_{e4}|^2 \lsim 10^{-6} - 10^{-7}$ for $170 \ {\rm MeV} \ < m_{\nu_4} < 450$
MeV \cite{na62_2018}; 

\item 
$|U_{\mu 4}|^2 \lsim 10^{-2}$ to $10^{-5}$ for $5 \ {\rm MeV} < m_{\nu_4} < 30$
MeV \cite{abela81}; 

\item 
$|U_{\mu 4}|^2 \lsim 10^{-4}$ for $3 \ {\rm MeV} \ < m_{\nu_4} < 19.5$ MeV 
\cite{daum87}; 

\item 
$|U_{\mu 4}|^2 \lsim 0.6 \times 10^{-5}$ for $16 \ {\rm MeV} < m_{\nu_4} < 29$
MeV and $|U_{\mu 4}|^2 \lsim 1 \times 10^{-5}$ for $29 \ {\rm MeV} < 
m_{\nu_4} < 32$ MeV \cite{triumf_pimu2}; 

\item 
$|U_{\mu 4}|^2 \lsim 10^{-8} - 10^{-9}$ for $200 \ {\rm MeV} \ <m_{\nu_4} < 
300$ MeV \cite{bnl949}; and,

\item 
$|U_{\mu 4}|^2 \lsim (1-4) \times 10^{-7}$ for $300 \ {\rm MeV} \ <m_{\nu_4} <
 450$ MeV \cite{na62_2018}. 

\end{itemize}
\noindent
Recently, the NA62 experiment at CERN  reported
more stringent preliminary upper limits on $|U_{e4}|^2$ and $|U_{\mu 4}|^2$:
\begin{itemize}

\item 
$|U_{e 4}|^2 \lsim (1 - 3) \times 10^{-9}$ for 
$150 \ {\rm MeV} \ < m_{\nu_4} < 400$ MeV, increasing to 
$|U_{e 4}|^2 \lsim (0.3 - 2) \times 10^{-8}$ for 
$400 \ {\rm MeV} < m_{\nu_4} < 450$ MeV, and 

\item

$|U_{\mu 4}|^2 \lsim ( 1 - 3) \times 10^{-8}$ for
$220 \ {\rm MeV} < m_{\nu_4} < 380$ MeV \cite{na62_kaon2019}. 

\end{itemize}
\noindent
Peak search experiments have also been conducted very near to the kinematic 
endpoint in 
$\pi^+ \to \mu^+ \nu_4$ decay, which occurs at $m_{\nu_4}=33.9122$ MeV 
\cite{daum1995,bryman_numao96,daum2000}. For $m_{\nu_4}=33.905$ MeV, a PSI 
experiment obtained an upper bound 
$BR(\pi^+ \to \mu^+ \nu_4) < 6.0 \times 10^{-10}$ (95 \% CL) \cite{daum2000}. 
From Eq. (\ref{usq_upper}), we estimate an upper limit 
\beqs
&& |U_{\mu 4}|^2 < 1.7 \times 10^{-8} \ (90 \% \ {\rm CL}) \ \ {\rm at} \ \ 
m_{\nu_4}=33.905 \ {\rm MeV}, \cr\cr
&&
\label{mnudaum2000}
\eeqs
which is shown in Fig. \ref{Umu4_figure}. 
An analysis of data on the $\mu$ capture reaction $\mu^- + {}^3{\rm He} \to 
\bar\nu_\mu + {}^3{\rm H}$ yielded upper limits on $|U_{\mu 4}|^2$ from 
$\sim 0.1$ to $\lsim 10^{-2}$ for $m_{\nu_4}$ in the interval from 62 MeV to 72
MeV \cite{deutsch83}. 
See \cite{pdg} for further limits and references to the literature). 

Upper limits on $|U_{e4}|^2$ vs. $m_{\nu 4}$ from $\pi_{e 2}$ and 
$K_{e 2}$  peak searches
are shown in Fig. ~\ref {Ue4_figure}, labeled as $\pi_{e2}$ PIENU,
$K_{e2}$ KEK, $K_{e2}$ NA62, and $K_{e2}$ NA62*, as well as the  
$B_{e2}$ limit presented in 
\cite{ul}, which will be discussed further below. Upper limits on 
$|U_{\mu 4}|^2$ vs. $m_{\nu 4}$ from $\pi_{\mu 2}$ and 
$K_{\mu 2}$  peak searches are shown in Fig.\ref {Umu4_figure}, labeled as 
$\pi_{\mu 2}$ PSI, $\pi_{\mu 2}$ PSI2, $\pi_{\mu 2}$ PIENU, $K_{\mu 2}$ KEK, 
$K_{\mu 2}$ BNL, $K_{\mu 2}$ NA62, and $K_{\mu2}$ NA62*. 

%

% =======================================================================

\begin{table}
  \caption{\footnotesize{Maximal values of the normalized kinematic rate 
      factor $\bar\rho(\delta_\ell^{(M)},\delta_{\nu_4}^{(M)})$ 
      for the two-body leptonic decay
      $M^+ \to \ell^+ \nu_4$ of the pseudoscalar meson $M^+$, where 
      $\ell=e, \ \mu$, together with the corresponding value of
      $m_{\nu_4}$, denoted $(m_{\nu_4})_{\bar\rho_{max}}$ (in MeV), 
      where this maximum is reached.}}
\begin{center}
\begin{tabular}{|c|c|c|} \hline\hline
Decay   &   $(m_{\nu_4})_{\bar\rho_{max}}$ & $\bar\rho_{max}$ \\
\hline\hline
$\pi^+ \to e^+ \nu_4$   & 80.6                & $1.105 \times 10^4$  \\
$K^+ \to e^+ \nu_4$     & 285                 & $1.38 \times 10^5$  \\
$D^+ \to e^+ \nu_4$     & $1.08 \times 10^3$  & $1.98 \times 10^6$  \\
$D_s^+ \to e^+ \nu_4$   & $1.14 \times 10^3$  & $2.20 \times 10^6$  \\
$B^+ \to e^+ \nu_4$     & $3.05 \times 10^3$  & $1.58 \times 10^7$  \\
\hline
$\pi^+ \to \mu^+ \nu_4$ & 3.46                & 1.00               \\
$K^+ \to \mu^+ \nu_4$   & 263                 & 4.13               \\
$D^+ \to \mu^+ \nu_4$   & $1.07 \times 10^3$  & 47.3               \\
$D_s^+ \to \mu^+ \nu_4$ & $1.13 \times 10^3$  & 52.4               \\
$B^+ \to \mu^+ \nu_4$   & $3.05 \times 10^3$  & 371                \\
\hline\hline
\end{tabular}
\end{center}
\label{rhobar_values}
\end{table}

% =====================================================================

\section{Constraints from Data on $e-\mu$ Universality}
\label{e_mu_universality_section}

% --------------------------------------------------------------------

\subsection{General Formalism}
\label{emuUGF}

In addition to producing a subdominant peak in the charged lepton momentum
$p_\ell$ at the value (\ref{pell}), the emission of a massive neutrino in the
two-body leptonic decay of a pseudoscalar meson $M^+$ would cause an apparent
deviation from the SM prediction for the ratio of decay rates or 
branching ratios,
\beq
R^{(M)}_{\ell/\ell'} \equiv 
\frac{BR(M^+ \to \ell^+\nu_\ell)}{BR(M^+ \to \ell'^+\nu_{\ell'})}  \ . 
\label{rm}
\eeq
The experimental measurements of $M^+ \to \ell^+ \nu_\ell$ 
include events with very soft photons; this is to be
understood implicitly below. 
By convention, we take $m_{\ell'} > m_{\ell}$. This deviation would constitute
an apparent violation of $e-\mu$ universality for the case $\ell=e$,
$\ell'=\mu$. In contrast to a peak search experiment with the decay $M^+ \to
\ell^+ \nu_\ell$, which places an upper bound on $|U_{\ell 4}|^2$ as a function
of $m_{\nu_4}$, a deviation in $R^{(M)}_{\ell/\ell'}$ depends, in general, on
both $|U_{ \ell 4}|^2$ and $|U_{\ell' 4}|^2$, as well as $m_{\nu_4}$ (see
Eqs. (\ref{rmgen}) and (\ref{rr} below). The non-observation of any additional
peak in the $dN/dp_\ell$ spectrum in two-body leptonic decays of $\pi^+$ and
$K^+$ was used via a retroactive data analysis in \cite{shrock80,shrock81a} and
in a series of dedicated peak-search experiments to set stringent upper limits
on $|U_{e4}|^2$ and $|U_{\mu 4}|^2$ (individually) as functions of
$m_{\nu_4}$. Furthermore, the non-observation of any deviation from $e-\mu$
universality in the ratio $R^{(M)}_{e/\mu}$ was used in
\cite{shrock81a,bryman83b,britton92} to obtain upper limits on lepton mixing,
as will be discussed further below. As was the case with peak search
experiments, in deriving a constraint from a comparison of a measured value of
$R^{(M)}_{\ell/\ell'}$ with the SM prediction for this ratio, one must take
into account that even if $m_{\nu_4}$ is small enough to be kinematically
allowed to occur in either or both of these decays, an experiment might reject
events involving emission of a $\nu_4$ if the momentum or energy of the
outgoing $\ell^+$ or $\ell'^+$ were below the cuts used in the event
reconstruction and data analysis.  We comment further on this below.

In Section \ref{peak_search_section} we  reviewed the general formalism 
describing effects of possible massive neutrino
emission in $M_{\ell 2}$ decays, i.e., the decays $M^+ \to \ell^+ \nu_\ell$,
where $\ell=e, \ \mu$, and, where allowed kinematically, also $\ell=\tau$
\cite{shrock80,shrock81a}.  Although the actual decays and branching ratios
depend on the pseudoscalar decay constants $f_M$ and the CKM mixing matrix
elements, these cancel in ratios of branching ratios, which can thus be
calculated to high precision and compared with experimental measurements. Let
us, then, consider the ratio of branching ratios (\ref{rm}).  In the Standard
Model, since the neutrino mass eigenstates $\nu_i$, $i=1,2,3$ have negligible
masses, this ratio is 
\beq
R^{(M)}_{\ell/\ell',SM} = \frac{m_\ell^2}{m_{\ell'}^2} \, \Bigg [ 
\frac{1-\frac{m_\ell^2}{m_M^2}}
     {1-\frac{m_{\ell'}^2}{m_M^2} } \Bigg ]^2 (1 + \delta_{RC}) \ , 
\label{rmsm}
\eeq
where $\delta_{RC}$ is the radiative correction 
\cite{earlyrad}-\cite{ml2radcor},
which takes into account soft photon emission, matching experimental
conditions. We define the ratio of the measured ratio of branching fractions,
$R^{(M)}_{\ell/\ell'}$ to the SM prediction for this ratio,
$R^{(M)}_{\ell/\ell',SM}$, as
\beq
\bar R^{(M)}_{\ell/\ell'} \equiv \frac{R^{(M)}_{\ell/\ell'}}
                                        {R^{(M)}_{\ell/\ell',SM}} \ .
\label{rmbar}
\eeq

Including the radiative correction $\delta_{RC}$, 
one has the following SM prediction for $R^{(\pi)}_{e/\mu,SM}$  
\cite{ms93}-\cite{sdb}
\beq
R^{(\pi)}_{e/\mu,SM} = (1.2352 \pm 0.0002) \times 10^{-4} \ . 
\label{rpi_emu_sm}
\eeq
The most recent and precise experimental measurement of this ratio of 
branching ratios was carried out by the PIENU experiment at TRIUMF, with the
result \cite{pienu2015} 
\beqs
R^{(\pi)}_{e/\mu} &=& (1.2344 \pm 0.0023_{stat} \pm 0.0019_{syst} ) 
\times 10^{-4} \ . \cr\cr
&&
\label{rpi_emu_triumf}
\eeqs
Combined with earlier data from lower-statistics experiments, this yields the
current weighted average listed by the Particle Data Group (PDG) for this
ratio, namely \cite{pdg}
\beqs
R^{(\pi)}_{e/\mu} &=& (1.2327 \pm 0.0023) \times 10^{-4} \ . \cr\cr
&&
\label{rpi_emu}
\eeqs
Using the PDG value of $R^{(\pi)}_{e/\mu}$, one finds 
\beq
\bar R^{(\pi)}_{e/\mu} = 0.9980 \pm 0.0019 \ . 
\label{rbar_pi_emu}
\eeq

For $R^{(K)}_{e/\mu}$, a similar analysis including radiative corrections 
\cite{ms93}-\cite{sdb} gives the SM prediction
\beq
R^{(K)}_{e/\mu,SM} = (2.477 \pm 0.001) \times 10^{-5}  \ . 
\label{rk_emu_sm}
\eeq
The current average experimental value which is dominated by the
measurement from the NA62 experiment \cite{na62_kemu}, is \cite{pdg} 
\beq
R^{(K)}_{e/\mu} = (2.488 \pm 0.009) \times 10^{-5}  \ . 
\label{rk_emu}
\eeq
To within the joint theoretical and experimental uncertainties, the measured
value of $R^{(K)}_{e/\mu}$ is in agreement with the SM prediction, as shown by
the ratio
\beq
\bar R^{(K)}_{e/\mu} = 1.0044 \pm 0.0037 \ . 
\label{rbar_k_emu}
\eeq

If a $\nu_4$ is emitted, then the ratio 
$R^{(M)}_{\ell/\ell',SM}$ changes to the following \cite{shrock80,shrock81a}:
\begin{widetext}
\beq
R^{(M)}_{\ell/\ell'}=
\Bigg [ \frac{[(1-|U_{\ell 4}|^2)\rho(\delta^{(M)}_\ell,0) + 
|U_{\ell 4}|^2 \rho(\delta^{(M)}_\ell,\delta^{(M)}_{\nu_4})}
{(1-|U_{\ell' 4}|^2)\rho(\delta^{(M)}_{\ell'},0) + 
|U_{\ell' 4}|^2 \rho(\delta^{(M)}_{\ell'},\delta^{(M)}_{\nu_4})} \Bigg ]
(1+\delta_{RC}) \ , 
\label{rmgen}
\eeq
\end{widetext}
where $\delta^{(M)}_\ell$ and $\delta^{(M)}_{\nu_4}$ were defined in Eq.
(\ref{deltas}).  In Eq. (\ref{rmgen}) we have used the fact that the leading
order radiative correction is independent of $m_{\nu_4}$ \cite{ml2radcor}.  As
noted above, in analyzing experimental data, one must take account of the fact
that unless an experiment is specifically searching for effects of possible
heavy neutrino emission, it would normally set cuts on the energy and/or
momentum of the outgoing charged lepton near to the value for the SM decay.  It
would thus reject events due to a sufficiently massive $\nu_4$ and would thus
measure an apparent total rate that would be reduced from the actual rate by
the factor $(1-|U_{e4}|^2)$.

Combining Eqs. (\ref{rmsm}) and (\ref{rmgen}), we have, for the ratio 
of branching ratios divided by the SM prediction, $\bar R^{(M)}_{\ell/\ell'}$,
\begin{widetext} 
\beq
\bar R^{(M)}_{\ell/\ell'} = \frac{ 1-|U_{\ell 4}|^2 + |U_{\ell 4}|^2 
\bar\rho(\delta^{(M)}_\ell,\delta^{(M)}_{\nu_4})}
{1-|U_{\ell' 4}|^2 + |U_{\ell' 4}|^2 
\bar\rho(\delta^{(M)}_{\ell'},\delta^{(M)}_{\nu_4}) } \ . 
\label{rr}
\eeq
\end{widetext}

With a given $M$, one can distinguish three different intervals for 
$m_{\nu_4}$:
\begin{itemize}
\item $I^{(M)}_1: \ m_{\nu_4} < m_M-m_{\ell'}$, 

\item $I^{(M)}_2: \ m_M-m_{\ell'} < m_{\nu_4} < m_M - m_\ell$, and

\item $I^{(M)}_3: \ m_{\nu_4} > m_M-m_\ell$.  

\end{itemize}
Thus,
(i) if $m_{\nu_4} \in I^{(M)}_1$, then both the $M^+ \to \ell^+ \nu_4$ and 
$M^+ \to \ell'^+ \nu_4$ decays can occur; 
(ii) if $m_{\nu_4} \in I^{(M)}_2$, then the $M^+ \to \ell^+ \nu_4$ can occur,
but the $M^+ \to \ell'^+ \nu_4$ decay is kinematically forbidden; and finally,
(iii) if $m_{\nu_4} \in I^{(M)}_3$, then both of the decays 
$M^+ \to \ell^+ \nu_4$ and $M^+ \to \ell'^+ \nu_4$ are kinematically 
forbidden. We recall the values of these intervals for the comparison of the
branching ratios for $M_{e2}$ and $M_{\mu 2}$ decays with $M = \pi^+$ and
$M=K^+$ (where we use the standard notation $M_{\ell 2}$ for the decay 
$M^+ \to \ell^+ \nu_\ell$).  Here, the mass intervals are 
\begin{itemize}

\item $I^{(\pi)}_1: \ m_{\nu_4} < 33.91$ MeV,

\item $I^{(\pi)}_2: \ 33.91 \ {\rm MeV} < m_{\nu_4} < 139.1$ MeV, 

\item $I^{(\pi)}_3: \ m_{\nu_4} > 139.1$ MeV. 

\item $I^{(K)}_1: \ m_{\nu_4} < 388.0$ MeV,

\item $I^{(K)}_2: \ 388.0 \ {\rm MeV} < m_{\nu_4} < 493.2$ MeV, and 

\item $I^{(K)}_3: \ m_{\nu_4} > 493.2 $ MeV. 
\end{itemize} 

The general forms of Eq. (\ref{rr}) are
\beqs
|U_{\ell 4}|^2 &<& \frac{\Big [1+|U_{\ell' 4}|^2
(\bar\rho(\delta^{(M}_{\ell'},\delta^{(M)}_{\nu_4})-1) \Big ]
\bar R^{(M)}_{\ell/\ell'}-1}
{\bar\rho(\delta^{(M}_\ell,\delta^{(M)}_{\nu_4})-1} \cr\cr
 && \quad {\rm for} \ m_{\nu_4} \in I^{(M)}_1 \  {\rm and} 
\label{uellsqbound_i1}
\eeqs
\beqs
\bar R^{(M)}_{\ell/\ell'} &=& \frac{ 1-|U_{\ell 4}|^2 + |U_{\ell 4}|^2
\bar\rho(\delta^{(M)}_\ell,\delta^{(M)}_{\nu_4})}
{1-|U_{\ell' 4}|^2 } \cr\cr
&& \quad {\rm for} \ m_{\nu_4} \in I^{(M)}_2 \ . 
\label{rr_bb2}
\eeqs
Consequently, for $m_{\nu_4} \in I^{(M)}_2$, from the upper limit on the
deviation of $BR(M^+ \to \ell^+ \nu_\ell)/BR(M^+ \to \ell'^+ \nu_{\ell'})$ from
its SM value, i.e., the upper limit on the deviation of 
$\bar R^{(M)}_{\ell/\ell'}$ from 1, an upper bound on $|U_{\ell 4}|^2$ can be
obtained.  Then,
\beqs
|U_{\ell 4}|^2 &<& \frac{(1-|U_{\ell' 4}|^2){\bar R}_{M;\ell/\ell'}-1}
{\bar\rho(\delta^{(M}_\ell,\delta^{(M)}_{\nu_4})-1} \quad {\rm for}
\ m_{\nu_4} \in I^{(M)}_2 \ . \cr\cr
&&
\label{uellsqbound_i2gen}
\eeqs
If $m_{\nu_4} \in I^{(M)}_3$, then Eq. (\ref{rr}) takes the still simpler form
\beq
\bar R^{(M)}_{\ell/\ell'} = \frac{1-|U_{\ell 4}|^2}{1-|U_{\ell' 4}|^2}
\quad {\rm for} \ m_{\nu_4} \in I^{(M)}_3 \ .
\label{rr_bb3}
\eeq
In general, if for a given $m_{\nu_4}$, one knows, e.g., from peak-search
experiments, that $|U_{\ell' 4}|^2$ is sufficiently small that the denominator
of (\ref{rr}) can be approximated well by 1, then an upper bound on the
deviation of $\bar R^{(M)}_{\ell/\ell'}$ from 1 yields an upper bound on
$|U_{\ell 4}|^2$:
\beq
|U_{\ell 4}|^2 < \frac{\bar R^{(M)}_{\ell/\ell'}-1}
{\bar\rho(\delta^{(M)}_\ell,\delta^{(M)}_{\nu_4})-1}.
\label{uellsqbound_i12}
\eeq
For cases in which $\ell=e$, this gives very stringent upper limits on
$|U_{\ell 4}|^2$ because $\bar\rho(\delta^{(M}_e,\delta^{(M)}_{\nu_4}) >> 1$
over much of the intervals $I^{(M)}_1$ and $I^{(M)}_2$, as can be seen from
Figs. 3-5 in \cite{shrock81a}.  For $m_{\nu_4} \in I^{(M)}_3$, the inequality
(\ref{uellsqbound_i12}) takes a simpler form, since
$\bar\rho(\delta^{(M)}_\ell,\delta^{(M)}_{\nu_4})=0$, namely
\beq
|U_{\ell 4}|^2 < 1-\bar R^{(M)}_{\ell/\ell'} \quad {\rm for} \ m_{\nu_4} \in 
I^{(M)}_3.
\label{uellsqbound_i3}
\eeq

We now apply this analysis to $R^{(\pi)}_{e/\mu}$, using
(\ref{uellsqbound_i12}) and (\ref{uellsqbound_i3}) with $M=\pi^+$, $\ell=e$,
and $\ell'=\mu$. From previous $\pi_{\mu 2}$ peak search experiments
\cite{abela81,daum87} and the recent \cite{triumf_pimu2}, and the calculation of
$\bar\rho(\delta^{(\pi)}_\mu,\delta^{(\pi)}_{\nu_4})$, it follows that $|U_{\mu
  4}|^2$ is sufficiently small for $m_{\nu_4} \in I^{(\pi)}_2$ that we can
approximate the denominator of Eq. (\ref{rr}) by 1.  From $\bar
R^{(\pi)}_{e/\mu}$ in Eq. (\ref{rbar_pi_emu}), using the procedure from
\cite{feldman_cousins}, we obtain the limit $\bar R^{(\pi)}_{e/\mu} <
1.0014$. Then, for $\nu_4 \in I^{(\pi)}_2$, we find
\beq
|U_{e 4}|^2 < \frac{\bar R^{(\pi)}_{e/\mu}-1}
{\bar\rho(\delta^{(\pi)}_e,\delta^{(\pi)}_{\nu_4})-1}
<  \frac{0.0014}{\bar\rho(\delta^{(\pi)}_e,\delta^{(\pi)}_{\nu_4})-1} .
\label{uellsqbound_i2pi}
\eeq
This bound is labeled as PIENU in Fig. \ref{Ue4_figure}.  If $m_{\nu_4} \in
I^{(\pi)}_3$, i.e., $m_{\nu_4} > 139$ MeV, then, using (\ref{uellsqbound_i3}),
we obtain the upper bound on $|U_{e4}|^2$ given by the flat line labeled
PIENU-H in Fig. \ref{Ue4_figure}.

We next obtain a bound on $|U_{e 4}|^2$ by applying the same type of analysis
to $R^{(K)}_{e/\mu}$. From $K_{\mu 2}$ peak search experiments
\cite{asano81,bnl949,na62_2018,na62_kaon2019} and the calculation of
$\bar\rho(\delta^{(K)}_\mu,\delta^{(K)}_{\nu_4})$, $|U_{\mu 4}|^2$ is
sufficiently small that we can approximate the denominator of Eq. (\ref{rr})
well by 1.  Using Eq. (\ref{rbar_k_emu}) for $\nu_4 \in I^{(K)}_2$, we find
\beq
|U_{e 4}|^2 < \frac{\bar R^{(K)}_{e/\mu}-1}
{\bar\rho(\delta^{(K)}_e,\delta^{(K)}_{\nu_4})-1}
<  \frac{0.010}{\bar\rho(\delta^{(K)}_e,\delta^{(K)}_{\nu_4})-1} .
\label{uellsqbound_i2k}
\eeq
This upper limit on $|U_{e4}|^2$ is labeled KENU in Fig. \ref{Ue4_figure}.  For
$m_{\nu_4} \in I^{(K)}_3$, i.e., $m_{\nu_4} > 493$ MeV, using
({\ref{uellsqbound_i3}), we obtain the flat upper bound labeled KENU-H in
Fig. \ref{Ue4_figure}. 

% ----------------------------------------------------------------------

\section{Bounds from Leptonic Decays of Heavy-Quark Mesons}
\label{heavy_meson_decay_section}

Two-body leptonic decays of heavy-quark pseudoscalar mesons
\cite{shrock80,shrock81a} are also valuable sources of information on sterile
neutrinos. We discuss the available bounds in this section.

% -------------------------------------------------------------------------

\subsection{Bounds from $D_s \to \ell^+ \nu_\ell$ Decays}

The two-body leptonic decays of the 
$D_s^+=(c\bar s)$ involve a large CKM mixing matrix factor $|V_{cs}|^2$.  
Two of these have been measured by the CLEO \cite{alexander09}, 
BABAR \cite{babar10}, Belle \cite{zupanc13}, and BES III 
\cite{ablikim16,ablikim19,bes3_pc} experiments, yielding the current values 
\beq
BR(D_s^+ \to \mu^+ \nu_\mu) = (5.49 \pm 0.17) \times 10^{-3} 
\label{br_ds_to_mu}
\eeq
and
\beq
BR(D_s^+ \to \tau^+ \nu_\tau)=(5.48 \pm 0.23) \times 10^{-2} \ . 
\label{br_ds_to_tau}
\eeq
Eq. (\ref{br_ds_to_mu}) is
a weighted average of CLEO, BABAR, Belle, and earlier BES III
measurements, combined with the most recent BES III result, $BR(D_s^+ \to \mu^+
\nu_\mu) = (5.49 \pm 0.16_{stat} \pm 0.15_{syst}) \times 10^{-3}$ (both
this new result and the weighted average (\ref{br_ds_to_mu}) are reported in
Ref. \cite{ablikim19}). 

Searches for $D_s^+ \to e^+ \nu_e$ have been carried out by CLEO 
\cite{alexander09}, BABAR \cite{babar10}, and Belle \cite{zupanc13}, giving 
the current upper bound 
\beq
BR(D_s^+ \to e^+ \nu_e) < 0.83 \times 10^{-4} \ . 
\label{br_ds_to_e}
\eeq
Hence, for the ratio of the $e$ and $\tau$ branching
ratios, one has the resultant upper limit 
\beq
R^{(D_s)}_{e/\tau} = \frac{BR(D_s^+ \to e^+ \nu_e)}
     {BR(D_s^+ \to \tau^+ \nu_\tau)}{}\Big |_{exp.} < 1.6 \times 10^{-3} \ . 
\label{brbr_dsedstau_exp}
\eeq
For this ratio, from \cite{ms93,ml2radcor} we calculate the radiative
correction $1+\delta_{RC}=0.948$. Substituting this in Eq. (\ref{rmsm}) with
$M=D_s$, $\ell=e$, and $\ell'=\tau$, we find that in the SM, this ratio of
branching ratios is
\beqs
R^{(D_s)}_{e/\tau,SM} &=& \frac{BR(D_s^+ \to e^+ \nu_e)_{SM}}
     {BR(D_s^+ \to \tau^+ \nu_\tau)_{SM}} \cr\cr
        &=& 2.29 \times 10^{-6} \ . 
\label{brbr_dsedstau_sm}
\eeqs
Hence, the current experimental upper limit on $BR(D_s^+ \to e^+ \nu_e)$ yields
the upper limit $\bar R^{(D_s)}_{e/\tau} < 7.0 \times 10^2$. 
Note that $m_{D_s} - m_\tau = 191$ MeV.

For $R^{(D_s)}_{e/\tau}$, the interval $I^{(D_s)}_2$ is $191 \ {\rm MeV} <
m_{\nu_4} < 1.457$ GeV.  We restrict $m_{\nu_4}$ to a lower-mass
subset of this full interval, for the following reason. In the $D_s^+ \to e^+
\nu_4$ decay, as $m_{\nu_4}$ increases from small values to its kinematic
limit, the momentum of the outgoing $e^+$ in the rest frame of the parent $D_s$
decreases from its SM value, $p_e \equiv |{\mathbf p}_e| = 0.984$ GeV. In order
for the event reconstruction procedure in a given experiment to count such a
decay as a $D_s^+ \to e^+ \nu_e$ decay, it is necessary that $p_e > p_{e,cut}$,
where $p_{e,cut}$ denotes a lower experimental cut on $p_e$.  A representative
value of this cut is the value $p_{e,cut}= 0.8$ GeV used in the BES III
experiment \cite{bes3_pc}.  The $e^+$ momentum decreases to $p_e=0.8$ GeV as
$m_{\nu_4}$ reaches the value $m_{\nu_4}=0.85$ GeV. Thus, we consider the
interval $0.191 \ {\rm GeV} \ < m_{\nu_4} < 0.85$ GeV.  For $m_{\nu_4}$ in this
interval, the ratio of
branching ratios of the observed $D_s^+ \to e^+ \nu_e$ and $D_s^+ \to \tau^+
\nu_\tau$ decays is given by Eq. (\ref{rr}) with $M = D_s^+$, $\ell=e$, and
$\ell'=\tau$. Hence, from Eq. (\ref{rr_bb2}), this ratio of branching ratios, 
divided by the value in the SM, is 
\beq
\bar R^{(D_s)}_{e/\tau} = \frac{1-|U_{e4}|^2 +
  |U_{e4}|^2\bar\rho(\delta_e^{(D_s)},\delta_{\nu_4}^{(D_s)})} 
{1-|U_{\tau 4}|^2} \ . 
\label{rr_dse_dstau}
\eeq
Requiring that the emission of the $\nu_4$ should not alter the experimentally
observed upper limit on $\bar R^{(D_s)}_{e/\tau}$ given above, we obtain the
following upper bound on $|U_{e4}|^2$ for $m_{\nu_4}$ in this mass range, which
is the special case of (\ref{uellsqbound_i2gen}) with $M=D_s$, $\ell=e$, and
$\ell'=\tau$:
\beq
|U_{e 4}|^2 < \frac{(1-|U_{\tau 4}|^2) \bar R^{(D_s)}_{e/\tau,ul}-1}
 {\bar\rho(\delta_e^{(D_s)},\delta_{\nu_4}^{(D_s)})-1} \ . 
\label{ue4sqlim_ds}
\eeq
This limit is largely independent of the $|U_{\tau 4}|^2$ term, since $|U_{\tau
  4}|^2$ is constrained to be less than upper bounds ranging from $\sim 0.1$ to
$\sim 0.01$ for $m_{\nu_4}$ in this mass range
\cite{gouvea,helo2011,batell2018}. For the minimal value of $m_{\nu_4}$ taken
here, namely $m_{\nu_4}=0.191$ GeV, the $\bar\rho$ function in
Eq. (\ref{ue4sqlim_ds}) is already quite large, having the value $1.37 \times
10^5$.  As $m_{\nu_4}$ increases to 0.85 GeV, this $\bar\rho$ function
increases to $1.83 \times 10^6$.  Thus, over this range of $m_{\nu_4}$, the
upper limit on $|U_{e 4}|^2$ in (\ref{ue4sqlim_ds}) decreases from $|U_{e 4}|^2
< 5.1 \times 10^{-3}$ to $|U_{e 4}|^2 < 3.8 \times 10^{-4}$.  We thus obtain
the upper bound on $|U_{e4}|^2$ labeled $D_{se2}$ in Fig. \ref{Ue4_figure}. In
the interval $450 \ {\rm MeV} < m_{\nu_4} < 850$ MeV, these upper bounds on
$|U_{e4}|^2$ (denoted as $D_{se2}$ in Fig. \ref{Ue4_figure}) are the best
available.  As was pointed out in \cite{ul}, dedicated peak-search experiments
to search for the heavy-neutrino decays $D_s^+ \to e^+ \nu_4$ and $D^+ \to e^+
\nu_4$ would be worthwhile and could improve our upper bound on $|U_{e4}|^2$.

In addition to the comparison of the branching ratios $BR(D_s^+ \to e^+ \nu_e)$
and $BR(D_s^+ \to \tau^+ \nu_\tau)$, it is also useful to comment on the
comparison of $BR(D_s^+\to\mu^+ \nu_\mu)$ and $BR(D_s^+ \to \tau^+\nu_\tau)$, 
both of which have been measured. From the experimental
results (\ref{br_ds_to_mu}) and (\ref{br_ds_to_tau}), the resultant
measured ratio of branching ratios is
\beq
R^{(D_s)}_{\mu/\tau} = 0.100 \pm 0.005 \ . 
\label{rr_ds_to_mu_over_tau}
\eeq
Substituting our calculated $1+\delta_{RC}=0.985$ for this decay in the general
formula (\ref{rmsm}), we find that the SM prediction for the branching ratio is
\beq
R^{(D_s)}_{\mu/\tau,SM} = 0.101 \ , 
\label{rr_ds_to_mu_over_tau_sm}
\eeq
so to this order, 
\beq
\bar R^{(D_s)}_{\mu/\tau}=0.990 \pm 0.05 \ .
\label{rr_dsmu_to_dstau}
\eeq
This yields the upper limit
$\bar R^{(D_s)}_{\mu/\tau} < \bar R^{(D_s)}_{\mu/\tau,ul}$, where
\beq
\bar R^{(D_s)}_{\mu/\tau,ul} = 1.05 \ . 
\label{rr_dsmustau_lim}
\eeq
Emission of a $\nu_4$ with non-negligible mass would change the ratio
(\ref{rr_dsmu_to_dstau}) to the expression in Eq. (\ref{rr}) with $M=D_s$,
$\ell=\mu$, and $\ell'=\tau$. The interval $I^{(D_s)}_2$ for this decay is $191
\ {\rm MeV} < m_{\nu_4} < 1.863$ GeV, and for $m_{\nu_4}$ in this interval,
Eq. (\ref{rr}) reduces to the expression in (\ref{rr_bb2}) with $M=D_s$,
$\ell=\mu$, and $\ell'=\tau$.  The maximum value of $m_{\nu_4}$ to enable a
large enough $p_\mu$ to satisfy an experimental lower momentum cut of 0.8 GeV
is $m_{\nu_4}=0.84$ GeV, which is almost the same as for the $D_s^+ \to e^+
\nu_4$ decay. We thus obtain an upper limit on $|U_{\mu 4}|^2$ which is the
special case of (\ref{uellsqbound_i2gen}) with $M=D_s$, $\ell=\mu$, and
$\ell'=\tau$, namely
\beq
|U_{\mu 4}|^2 < \frac{(1-|U_{\tau 4}|^2) \bar R^{(D_s)}_{\mu/\tau,ul}-1}
 {\bar\rho(\delta_\mu^{(D_s)},\delta_{\nu_4}^{(D_s)})-1} \ . 
\label{umu4sqlim_ds}
\eeq
Given that $|U_{\tau 4}|^2 << 1$, this reduces to the special case of
Eq. (\ref{uellsqbound_i12}) with $M=D_s$, $\ell=\mu$, and $\ell'=\tau$.  For
$m_{\nu_4}=0.191$ GeV, the $\bar\rho$ function in Eq. (\ref{umu4sqlim_ds}) has
the value 4.22. As $m_{\nu_4}$ increases to 0.84 GeV, this $\bar\rho$ function
increases to 43.4.  With $|U_{\tau 4}|^2 << 1$, the resultant upper bound on
$|U_{\mu 4}|^2$ is shown in Fig. \ref{Umu4_figure}.  This bound decreases from
$\sim 10^{-2}$ to $\sim 10^{-3}$ over this range of $m_{\nu_4}$. In the lower
part of this interval, $0.22 \ {\rm GeV} < m_{\nu_4} < 0.38 \ {\rm GeV}$, the
BNL E949 and NA62 peak search experiments with $K_{\mu 2}$ decay have set 
more stringent upper bounds, but
in the upper part of the interval, between $m_{\nu_4} = 0.46$ GeV and 
$m_{\nu_4}=0.84$ GeV, our upper bound
on $|U_{\mu 4}|^2$ from this analysis of $D_s$ decays is the best current
direct laboratory upper bound.

% --------------------------------------------------------------------

\subsection{Bounds from $D^+ \to \ell^+ \nu_\ell$ Leptonic Decays} 

In the case of the $D^+$ meson, the $c \bar d$ annihilation amplitude is
suppressed by the CKM factor $|V_{cd}|^2$ relative to semileptonic and
hadronic decay channels, which can proceed by $c \to s$ charged-current
vertices and hence involve the much larger $|V_{cs}|^2$ factor in the
rates. There is significant phase-space suppression of the $D^+ \to \tau^+
\nu_\tau$ channel, since $m_{D^+}-m_\tau$ is only 92.8 MeV.  For one of these
leptonic $D$ decays, one has an upper limit, namely 
\beq
BR(D^+ \to e^+ \nu_e) < 0.88 \times 10^{-5} \ . 
\label{br_d_to_e}
\eeq
The branching ratio for 
$D^+ \to \mu^+ \nu_\mu$ has been measured by CLEO and BES III 
\cite{eisenstein08,ablikim14,pdg} as 
\beq
BR(D^+ \to \mu^+ \nu_\mu) = (3.74 \pm 0.17) \times 10^{-4} \ . 
\label{br_d_to_mu}
\eeq
Recently, BES III has measured the branching ratio for $D^+ \to \tau^+
\nu_\tau$ \cite{bes3_Dtaunu} as 
\beq
BR(D^+ \to \tau^+ \nu_\tau) =  (1.20 \pm 0.24_{\rm stat.} 
\pm 0.12_{\rm syst.} ) \times 10^{-3} \ . 
\label{br_d_to_tau_val}
\eeq
With the radiative correction $1+\delta_{RC}=0.963$, the SM 
prediction for the ratio of these branching ratios is, from Eq. (\ref{rm}), 
\beq
\bar R^{(D)}_{e/\mu,SM}{}\Big |_{SM} = 2.27 \times 10^{-5} \ . 
\label{brbr_dedtau_sm}
\eeq
From the experimental limit (\ref{br_d_to_e}) and measurement 
(\ref{br_d_to_mu}), we have the 90 \% CL upper limit 
\beq
\bar R^{(D)}_{e/\mu}{}\Big |_{exp} < 2.5 \times 10^{-2} \ . 
\label{rr_d_to_emu_lim}
\eeq
With $\nu_4$ emission, this ratio would be changed to 
\beq
\bar R^{(D)}_{e/\mu} = \frac{1-|U_{e4}|^2 + 
|U_{e4}|^2 \bar\rho(\delta_e^{(D)},\delta_{\nu_4}^{(D)})}
{1-|U_{\mu 4}|^2+|U_{\mu4}|^2\bar\rho(\delta_\mu^{(D)},\delta_{\nu_4}^{(D)})} 
\ . 
\label{rr_d_to_emu}
\eeq
Requiring that $\bar R^{(D)}_{e/\mu}$ not violate the upper bound 
(\ref{rr_d_to_emu_lim}) yields correlated upper limits on $|U_{e4}|^2$ and 
$|U_{\mu 4}|^2$ as a function of $m_{\nu_4}$. 

% ---------------------------------------------------------------------

\subsection{Bounds from $B^+ \to \ell^+ \nu_\ell$ Leptonic Decays} 

Here we analyze constraints from two-body leptonic $B^+$ decays. These decays
involve $u \bar b$ annihilation and hence are suppressed by the small CKM
factor $|V_{ub}|^2$ relative to semileptonic and hadronic $B^+$ decays
involving the larger CKM factor $|V_{cb}|^2$.  Currently, there is an upper
limit on one leptonic $B^+$ decay,
\beq
BR(B^+ \to e^+\nu_e) < 0.98 \times 10^{-6}
\label{br_benu}
\eeq
from Belle \cite{satoyama07} and BABAR \cite{aubert09},
and measurements of the other two, namely
\beqs
&& BR(B^+ \to \mu^+ \nu_\mu) = (6.46 \pm 2.22_{stat} \pm 1.60_{syst} )
\times 10^{-7} \cr\cr
&&
\label{br_bmunu}
\eeqs
from Belle \cite{belle_bmunu2018}, 
\beqs
BR(B^+ \to \mu^+ \nu_\mu) = (5.3 \pm 2.0_{stat} \pm 0.9_{syst}) 
\times 10^{-7}  \cr\cr
&&
\label{br_bmunu_moriond}
\eeqs
from a Belle update \cite{belle_moriond2019,moriond2019}, and
\beq
BR(B^+ \to \tau^+ \nu_\tau) = (1.09 \pm 0.24) \times 10^{-4}
\label{br_btaunu}
\eeq
from BABAR \cite{lees13} and Belle \cite{hara13,kronenbitter15}.
Both the published and preliminary updated values of the $BR(B^+ \to
\mu^+\nu_\mu)$ are in agreement with the SM prediction \cite{belle_bmunu2018}
\beq
BR(B^+ \to \mu^+ \nu_\mu)_{SM} = (3.80 \pm 0.31) \times 10^{-7} \ .
\label{br_bmunu_sm}
\eeq
The measured value of $BR(B^+ \to \tau^+ \nu_\tau)$ in
(\ref{br_btaunu}) is also in agreement with the
SM prediction \cite{kronenbitter15,ckmfitter}
\beq
BR(B^+ \to \tau^+ \nu_\tau)_{SM} = (0.75^{+0.10}_{-0.05}) \times 10^{-4} \ . 
\label{b_to_tau_sm}
\eeq
From \cite{ms93} we calculate the radiative correction
factor $1+\delta_{RC}= 0.942$ for $R^{(B)}_{e/\tau,SM}$.
Combining this with (\ref{br_btaunu}) and (\ref{rmsm}), we then obtain the
SM prediction
\beq
BR(B^+ \to e^+ \nu_e)_{SM} = (1.08 \pm 0.24) \times 10^{-11} \ .
\label{br_benu_sm}
\eeq

A recent experiment to search for $B^+ \to e^+ X^0$ and $B^+ \to \mu^+ X^0$ was
carried out by Belle \cite{park2016}, where $X^0$ is a weakly interacting
particle that does not decay in the detector. Assuming that $X^0 = \nu_4$, one
can use the results of this experiment to set upper limits on $|U_{e 4}|^2$ and
$|U_{\mu 4}|^2$. For $m_{\nu_4}$ in the range from 0.1 GeV to 1.4 GeV, this
experiment obtained an upper limit on $BR(B^+ \to e^+ \nu_4)$ of $2.5 \times
10^{-6}$, while in the interval of $m_{\nu_4}$ from 1.4 GeV to 1.8 GeV, this
upper limit increased to $7 \times 10^{-6}$. In the range of $m_{\nu_4}$ from
0.1 to 1.3 GeV, the experiment obtained (non-monotonic) upper limits on $BR(B^+
\to \mu^+ \nu_4)$ of approximately $2 \times 10^{-6}$ to $4 \times 10^{-6}$,
and in the interval of $m_{\nu_4}$ from 1.3 GeV to 1.8 GeV, it obtained upper
limits varying from $2 \times 10^{-6}$ to $1.1 \times 10^{-5}$.  These limits
are less restrictive than the bounds (\ref{br_benu}) and (\ref{br_bmunu}), but
have the advantage of being reported for specific values of $m_{\nu_4}$.

Substituting the experimental upper limit on $BR(B^+ \to e^+ \nu_4)$ as a
function of $m_{\nu_4}$ from \cite{park2016} into the relevant special case of
(\ref{usq_upper_full}) with $M = B^+$ and $\ell = e$, we obtain the upper
bound on $|U_{e 4}|^2$ as a function of $m_{\nu_4}$ shown in
Fig. \ref{Ue4_figure}. This upper bound decreases from 0.83 to $3.4 \times
10^{-2}$ as $m_{\nu_4}$ increases from 0.1 GeV to 1.2 GeV. Since the
experimental upper limit on $BR(B^+ \to e^+ \nu_4)$ is less stringent as
$m_{\nu_4}$ increases from 1.4 to 1.8 GeV, the same is true of the resultant
upper limit on $|U_{e4}|^2$; for example, if $m_{\nu_4}=1.6$ GeV, we get
$|U_{e4}|^2 < 5.4 \times 10^{-2}$.

Carrying out the analogous procedure with the upper bound on $BR(B^+ \to \mu^+
\nu_4)$ from \cite{park2016}, we obtain an upper limit on $|U_{\mu 4}|^2$ that
decreases from 0.83 to $3.4 \times 10^{-2}$ as $m_{\nu}$ increases from 0.1 GeV
to 1.2 GeV. As $m_{\nu_4}$ increases from 1.2 to 1.5 GeV and then to 1.8 GeV,
the upper limit on $BR(B^+ \to \mu^+ \nu_4)$ from \cite{park2016} rises from
approximately $3 \times 10^{-6}$ to $1.1 \times 10^{-5}$ and then decreases
again to $3 \times 10^{-6}$.  In this interval of $m_{\nu_4}$ masses, using the
appropriate special case of (\ref{usq_upper_full}), we obtain upper limits
ranging from $|U_{\mu 4}|^2$ of 0.12 for $m_{\nu_4}=1.5$ GeV to $|U_{\mu 4}|^2$
of $2.7 \times 10^{-2}$ at $m_{\nu_4}=1.8$ GeV.  See also \cite{batell2018}.
Further peak searches for $B^+ \to e^+ \nu_4$ and $B^+ \to \mu^+ \nu_4$ with
Belle II would be valuable and could improve the limits from
Ref. \cite{park2016}. Moreover, when measurements of two-body leptonic decays
of $B_c^+$ mesons become available, it would also be of interest to use them to
constrain lepton mixing matrix coefficients.

As was true for the other decays, in obtaining these limits from leptonic $B$
decays, it is assumed that the only new physics is the emission of the massive
$\nu_4$.  However, in the $B$ system there are currently 
several quantities whose experimental measurements are in possible
tension with SM predictions, including, for example, the ratios of branching
ratios $R(D^{(*)}) = BR(B \to D^{(*)} \tau \bar\nu_\tau)/ BR(B \to D^{(*)} \ell
\bar\nu_\ell)$, where $\ell=e, \ \mu$, and the ratio $R(K^{(*)}) = BR(B \to
K^{(*)}e^+e^-)/BR(B\to K^{(*)} \mu^+\mu^-)$ (see, e.g., \cite{cerri,forti}).

% ***======================================================================

\section{Constraints from $\mu$ Decay}
\label{mu_decay_section}

% ---------------------------------------------------------------------

\subsection{General Analysis with Massive Neutrino Emission}
\label{munu_subsection}

In this section we discuss constraints from $\mu$ decays.
The lifetime of the $\mu^+$ was measured to 0.5 ppm accuracy by the MuLan
experiment at PSI \cite{mulan}, yielding the value $G_F = 1.1663787(6)
\times 10^{-5}$ GeV$^{-2}$ with the implicit assumption of decays only
into the three known neutrino mass eigenstates.  With this assumption,
the uncertainty in this determination of $G_F$ is mainly from the
experimental measurement; it is estimated that the uncertainty due
to radiative corrections \cite{ks,ritbergen_stuart_mu,czarnecki_mu} is 
approximately 0.14 ppm and the uncertainty from the measured value of 
$m_\mu$ is 0.08 ppm \cite{mulan}.

However, as was pointed out and analyzed in \cite{shrock80,shrock81b}, 
in the presence of neutrino masses and lepton mixing, the
decay $\mu \to \nu_\mu e \bar\nu_e$ would actually consist of the decays
$\mu \to \nu_i e \bar\nu_j$ into the individual
mass eigenstates $\nu_i$ and $\bar\nu_j$ in the interaction eigenstates
$\nu_\mu$ and $\bar\nu_e$, where $1 \le i,j \le
3+n_s$, as allowed by phase space. The emission of massive neutrino(s)
with non-negligible mass(es) in muon decay would produce several
changes relative to the Standard Model.  These include
(i) kink(s) in the observed
electron energy spectrum associated with the fact that the maximum
electron energy in the rest frame of the parent $\mu$ is reduced from its 
SM value with neutrinos of negligibly small masses,
\beq
E_{e,max} = \frac{m_\mu^2+m_e^2}{2m_\mu}
\label{emax_mu_sm}
\eeq
to
\beq
E_{e,max,ij} = \frac{m_\mu^2+m_e^2-(m_{\nu_i} + m_{\nu_j})^2}{2m_\mu} \ ;
\label{emax_mu_sm_nu}
\eeq
(ii) reduction of the differential and total decay rate; (iii) a reduction in
the apparent value of the Fermi coupling $G_F$, relative to its value in the
Standard Model with neutrinos of negligibly small masses; and (iv) changes in
the spectral parameters $\rho$ and $\eta$, and, for a polarized muon, $\xi$,
and $\delta$, that have been used to fit the differential decay spectrum of the
muon.  Ref. \cite{shrock81b} calculated the changes in these spectral
parameters that would be caused by emission of a massive (anti)neutrino in
$\mu$ decay and used existing data to set upper limits on lepton mixing
coefficients as functions of neutrino mass. From data on the $\rho$ parameter
describing the $e^+$ momentum distribution in unpolarized $\mu^+$ decay,
Ref. \cite{shrock81b} derived an upper limit on $|U_{r4}|^2$, where $r=e, \
\mu$ in the interval $m_{\nu_4}$ up to 70 MeV, extending down to a few times
$10^{-3}$ at $m_{\nu_4}=30$ MeV.  This constraint applies to both $|U_{e4}|^2$
and $|U_{\mu 4}|^2$ since the $\nu_4$ or $\bar\nu_4$ can be emitted at either
the charged-current vertex with the $\mu$ or with the $e$.  The upper bound on
$|U_{e4}|^2$ from $\mu$ decay is not as restrictive as upper bounds from
$\pi_{e2}$ or $K_{e2}$ decay. However, the upper bound on $|U_{\mu 4}|^2$ from
$\mu$ decay is valuable for an interval of $m_{\nu_4}$ that is not covered by
peak search experiments, namely the interval above the kinematic endpoint for
$\pi_{\mu 2}$ decay at $m_{\nu_4}=33.9$ MeV and below the value of $m_{\nu_4}
\simeq 40$ MeV, which was the lowest value at which a $K_{\mu 2}$ peak search
experiment (at KEK \cite{nordkirchen}) obtained an upper limit on $|U_{\mu
  4}|^2$.  In \cite{shrock_vpi,shrock_snowmass82} it was pointed out that
because, in the presence of massive neutrino emission in $\mu$ decay, the value
of $G_{F,app}$ extracted in the framework of the SM is smaller than the true
value of $G_F$, this would lead to predictions of the masses of the $W$ and
$Z$, that would be larger than the true values, and these effects were
calculated. Subsequent discussions of massive neutrino effects in $\mu$ decay
include \cite{gninenko}, \cite{kalyniak_ng82}, \cite{dixit_kalyniak_ng},
\cite{bryman_picciotto}, and \cite{batell2018}. In particular, the TWIST
experiment at TRIUMF measured $\rho$ with greater accuracy \cite{twist}. Using
an analysis similar to that in \cite{shrock81b} applied to the TWIST data, one
obtains upper limits on $|U_{\mu4}|^2$ extending down to $2 \times 10^{-3}$ at
$m_{\nu_4}=30$ MeV (e.g., \cite{batell2018}).

Let us consider the change in the total rate as a consequence of
muon decays to a neutrino mass eigenstate $\nu_4$ with a
non-negligible mass. In the SM with neutrinos of negligibly
small mass, the rate for $\mu$ decay has the form
\beq
\Gamma_{\mu,SM} = \frac{G_F^2 m_\mu^5}{192\pi^3} \, (1+\delta_\alpha) \, \bar
\Gamma_{\mu,SM} \ . 
\label{gamma_mu_sm}
\eeq
Here we have separated off a rate factor 
\beq
\bar \Gamma_{\mu,SM} = f(a_e^{(\mu)},0,0) \ , 
\label{gammabar_mu_sm}
\eeq
where $f$ is a dimensionless kinematic function resulting from the integration
over the three-body final-state phase space, which depends on three arguments,
namely the (squares of the) ratios of each of the final-state 
particle masses to the muon mass.
Finally, in Eq. (\ref{gamma_mu_sm}), the $\delta_\alpha$ term incorporates
electroweak corrections and has the leading-order value
$\delta_\alpha=-[\alpha_{em}/(2\pi)][\pi^2-(25/4)] = -4.2 \times 10^{-3}$
\cite{ks}.  For the SM with neutrinos of negligibly small mass, the kinematic
function $f$ is 
\beq
f(a,0,0) = (1-8a+a^2)(1-a^2) + 12a^2 \ln \bigg ( \frac{1}{a} \bigg )
\label{fsm}
\eeq
with
\beq
a_e^{(\mu)} = \frac{m_e^2}{m_\mu^2} = 2.339010 \times 10^{-5} \ .
\label{ae}
\eeq
Numerically, $f(a_e^{(\mu)},0,0) = 1-(1.87 \times 10^{-4})$.  The SM kinematic
function for $\mu$ decay has the series expansion 
\beq
f(a,0,0) = 1-8a + O(\{ a^2, \ a^2\ln a \})
\label{fsm_taylor}
\eeq
with $a=a^{(\mu)}_e$.  Because $a^{(\mu)}_e << 1$, $f(a_e^{(\mu)},0,0)$ is 
very well approximated, to three-figure accuracy, by the first two terms
in its series expansion, $1-8a_e^{(\mu)}$.

For our case, from the general formulas in \cite{shrock81b},
the $\mu$ decay rate is given by 
\begin{widetext}
  \beqs
  \bar\Gamma_\mu &=& (1-|U_{e4}|^2)(1-|U_{\mu 4}|^2)f(a_e^{(\mu)},0,0) +
    (1-|U_{e4}|^2)|U_{\mu 4}|^2 f(a_e^{(\mu)},0,a_{\nu_4}^{(\mu)})
    \cr\cr
    &+& |U_{e4}|^2(1-|U_{\mu 4}|^2) f(a_e^{(\mu)},a_{\nu_4}^{(\mu)},0) +
            |U_{e4}|^2|U_{\mu 4}|^2
            f(a_e^{(\mu)},a_{\nu_4}^{(\mu)},a_{\nu_4}^{(\mu)}) \cr\cr
   &=& \Gamma_{\mu,SM} \bigg [ (1-|U_{e4}|^2)(1-|U_{\mu 4}|^2) +
    (1-|U_{e4}|^2)|U_{\mu 4}|^2 \bar f(a_e^{(\mu)},0,a_{\nu_4}^{(\mu)})
    \cr\cr
    &+& |U_{e4}|^2(1-|U_{\mu 4}|^2) \bar f(a_e^{(\mu)},a_{\nu_4}^{(\mu)},0) +
|U_{e4}|^2|U_{\mu 4}|^2 \bar f(a_e^{(\mu)},a_{\nu_4}^{(\mu)},a_{\nu_4}^{(\mu)})
 \bigg ] \ ,
  \label{gamma_mu}
  \eeqs
\end{widetext}
where $\bar f(x,y,z))$ is the ratio of the kinematic phase space
integral for each of the decays divided by the kinematic integral for the SM
decay (\ref{fsm}):
\beq 
\bar f(x,y,z)=\frac{f(x,y,z)}{f(x,0,0)}
\label{rho}
\eeq
with 
\beq
a_{\nu_4}^{(\mu)} = \frac{m_{\nu_4}^2} {m_\mu^2} \ .
\label{anu}
\eeq
Here and below, the kinematic function $f(x,y,z)=0$ if the decay is
kinematically forbidden, i.e., if $\sqrt{x}+\sqrt{y}+\sqrt{z} \ge 1$.
The four terms in Eq. (\ref{gamma_mu}) arise from the decays (a) $\mu
\to \nu_i e \bar\nu_j$; (b) $\mu \to \nu_4 e \bar\nu_i$; (c) $\mu \to
\nu_i e \bar\nu_4$; and (d) $\mu \to \nu_4 e \bar\nu_4$, where here
$\nu_i$ and $\nu_j$ denote the known three neutrino mass eigenstates,
whose masses are negligibly small in $\mu$ decay.  Note that the
second and third terms are present only if $m_\mu > m_e + \nu_4$, and
the fourth term is present only if $m_\mu > m_e +
2m_{\nu_4}$. Furthermore, the fourth term is strongly suppressed
because it involves the product of the squares of two small leptonic
mixing matrix coefficients, $|U_{e4}|^2|U_{\mu 4}|^2$, and because of
the smaller phase space if $m_{\nu_4}/m_\mu$ is substantial.  Hence,
to evaluate Eq. (\ref{gamma_mu}) for $\bar\Gamma_\mu$, to a very good
approximation, we may drop the last term, and hence we need only the
kinematic function $f(x,y,0)$, which was calculated in 
Ref. \cite{shrock81b}.  A basic symmetry property of the kinematic
function is that \cite{shrock81b}
\beq
f(x,y,z)=f(x,z,y) \ , 
\label{fsym}
\eeq 
so the second and third terms in Eq. (\ref{gamma_mu}) have the same
kinematic factor, $\bar f(a_e^{(\mu)},0,a_{\nu_4}^{(\mu)})=
\bar f(a_e^{(\mu)},a_{\nu_4}^{(\mu)},0)$.

The apparent value of the Fermi coupling, $G_{F,app}$, obtained from
the measurement of the $\mu$ decay rate is given by
\beq
\frac{G_{F,app}^2}{G_F^2} = \frac{\Gamma_\mu}{\Gamma_{\mu,SM}} \equiv \kappa
\label{gfrel}
\eeq
and is less than unity if (anti)neutrinos with non-negligible masses
are emitted in $\mu$ decay \cite{shrock81b}. Explicitly,
  \begin{widetext}
  \beqs
  \frac{G_{F,app}^2}{G_F^2} &=& (1-|U_{e4}|^2)(1-|U_{\mu 4}|^2) +
    (1-|U_{e4}|^2)|U_{\mu 4}|^2 \bar f(a_e^{(\mu)},0,a_{\nu_4}^{(\mu)})
    \cr\cr
    &+& |U_{e4}|^2(1-|U_{\mu 4}|^2) \bar f(a_e^{(\mu)},a_{\nu_4}^{(\mu)},0) +
|U_{e4}|^2|U_{\mu 4}|^2\bar f(a_e^{(\mu)},a_{\nu_4}^{(\mu)},a_{\nu_4}^{(\mu)}) 
\ . \cr\cr
  &&
\label{gfratio}
\eeqs
\end{widetext}
In the SM, the predicted mass of the $Z$ is determined in terms of
$\alpha=e^2/(4\pi)$, the weak mixing angle $\theta_W = \arctan(g'/g)$, and 
$G_{F,app}$ by
\beq
m_{Z,pred} = \bigg ( \frac{\pi \alpha}{2^{1/2} G_{F,app}} \bigg )^{1/2} 
\frac{1}{\sin\theta_W \cos\theta_W} \, (1+\delta_{Z,RC}) \ , 
\label{mzpred}
\eeq
and $m_{W,pred}=m_{Z,pred} \cos\theta_W$, where $\delta_{Z,RC}$ is the
radiative correction \cite{msv80}.  As pointed out in 
\cite{shrock_vpi,shrock_snowmass82},
in the presence of massive neutrino emission in
$\mu$ decay, these predicted values of $m_Z$ and $m_W$ would be larger than
the true values, since $G_{F,app} < G_F$:
\beq
m_{Z,true} = \kappa^{1/2} m_{Z,pred} < m_{Z,pred}
\label{mz_true_vs_pred}
\eeq
and
\beq
m_{W,true} = \kappa^{1/2} m_{W,pred} < m_{W,pred} \ . 
\label{mw_true_vs_pred}
\eeq
The effects on the $W$ and $Z$ widths were also discussed in
\cite{shrock_vpi,shrock_snowmass82}.  The agreement between the predicted
and observed masses and widths of the $W$ and $Z$ thus yield constraints
on leptonic mixing angles as functions of $m_{\nu_4}$.  With current 
values of $m_W$, $m_Z$, $\Gamma_W$, and $\Gamma_Z$, these
imply $|U_{\ell 4}|^2 \lsim 10^{-2}$ (e.g., \cite{batell2018}).

As mentioned above, the test of relative agreement of ${\cal F}t$ values
obtained from the set of 14 superallowed nuclear beta decays in
\cite{hardy_towner2015,hardy_towner2018} is independent of $G_{F,app}$
since this divides out in the ratios of the ${\cal F}t$ values.
However, depending on $m_{\nu_4}$, $|U_{e4}|^2$, and $|U_{\mu 4}|^2$,
the result would generically be that the value of $|V_{ud}|$ obtained
from these nuclear beta decays would not be equal to the true value,
because of both the reduction of the rates for the various nuclear
beta decays and the fact that the value of $G_{F,app}$ used in
Eq. (\ref{nucldecrate}) would be different from the true value. In
turn, this would generically lead to a spurious apparent violation of
the first-row CKM unitarity test.  Whether the apparent value of
$|V_{ud}|$ would be larger or smaller than the true value would depend
on the values of $m_{\nu_4}$, $|U_{e4}|^2$, and $|U_{\mu 4}|^2$ and
thus on the relative effects of massive neutrino emission in muon
decay and in the nuclear beta decays used to obtain $|V_{ud}|$.

Since the determination of $|V_{ud}|$ from the superallowed nuclear beta decays
depends on the input value of $G_{F,app}$ from muon decay, an apparent
violation of the first-row CKM unitarity relation $\Sigma=1$ could indicate the
presence of effects of new physics beyond the Standard Model (BSM) in muon
decay.  Although our discussion above has focused on the effect of the possible
emission of neutrino(s) of non-neglible masses and couplings in muon decay, we
note that there could also be exotic muon decays in BSM scenarios that would
appear observationally to be the same as $\mu^+ \to \bar\nu_\mu e^+ \nu_e$,
i.e., $\mu^+ \to e^+ + {\rm missing \ neutrals}$, where the additional neutral
particles are weakly interacting. An explicit example studied in the context of
supersymmetric extensions of the SM was the decay $\mu^+ \to e^+ \tilde\gamma
\tilde\gamma$, where $\tilde\gamma$ denotes the photino \cite{megg}. An
analogous decay involving hadrons was $K^+ \to \pi^+ \tilde\gamma \tilde\gamma$
\cite{kpigg}, which would appear observationally as $K^+ \to \pi^+ + {\rm
  missing \ neutrals}$ and hence would be experimentally indistinguishable from
the SM decay $K^+ \to \pi^+ \nu\bar\nu$ \cite{neutralinos}. (In modern
notation, these decays would be denoted as $\mu^+ \to e^+ \tilde\chi^0
\tilde\chi^0$ and $K^+ \to \pi^+ \tilde\chi^0 \tilde\chi^0$, where
$\tilde\chi^0$ is a neutralino.)  As was noted in \cite{megg}, the existence of
the decay $\mu^+ \to e^+ \tilde\gamma \tilde\gamma$ by itself would lead to an
apparent value of $G_{F,app}$ larger than the true value, opposite to the
effect of massive neutrino emission.  Another possibility for an exotic $\mu$
decay is $\mu \to e + x$, where $x$ is a neutral, light, weakly interacting
boson; upper limits on this were given in \cite{bryman_clifford}
\cite{TWIST_mux}.  Another
example of this type of additional exotic $\mu$ decay was studied in a model
with dynamical electroweak symmetry breaking \cite{etc}, in which the $\mu^+
\to e^+$ transition would be mediated by a neutral virtual massive
generation-changing vector boson, which then would produce a final-state
$\bar\nu_\mu \nu_e$ pair (see also \cite{blucher_marciano}).

% ---------------------------------------------------------------------------

\subsection{Limit on Exotic $\mu$ Decay Modes} 
\label{exotic_subsection}

If there are no light sterile neutrinos relevant for $\mu$ decay, but there are
additional exotic muon decays such as in the examples above, then, since the
experimentally extracted value of $G_{F,app}$ would be larger than the true
$G_F$, the resultant apparent value of $|V_{ud}|$ obtained from the
superallowed nuclear beta decays, denoted $|V_{ud}'|$, would be smaller than
the true value.  In turn, this would yield an apparent spurious violation of
CKM unitarity in which the apparent value of $\Sigma$ would be less than unity.
Since an exotic BSM decay channel would increase $\Gamma_\mu$ relative to the
SM value $\Gamma_{\mu,SM}$, while emission of heavy neutrino(s) would decrease
$\Gamma_\mu$ relative to $\Gamma_{\mu,SM}$, it is possible, in principle, for
both of these non-SM effects to be present and to tend to cancel each other,
yielding a resultant $\Gamma_\mu$ close to $\Gamma_{\mu,SM}$.  However, in the
absence of any symmetry reason, such a cancellation may be regarded as
unlikely.  Accordingly, in our analyses, we will treat each of these two cases
individually.

If one considers the possibility that no heavy sterile
(anti)neutrinos are emitted in $\mu$ decay but instead, there is an
exotic extra decay channel (indicated with subscript $ext$)
with rate $\Gamma_{\mu,ext}$, then
the total decay rate would be $\Gamma_\mu = \Gamma_{\mu,SM} +
\Gamma_{\mu,ext}$. Let us denote $\Gamma_{\mu,SM} \equiv G_F^2
\hat{\Gamma}_{\mu,SM}$ and the branching ratio of the exotic decay mode
as $BR_{\mu,ext}=\Gamma_{\mu,ext}/\Gamma_\mu$. 
Experimentalists would then extract the apparent value $G_{F,app}$ as
\beqs
G_{F,app}^2 \hat{\Gamma}_{\mu,SM} &=& \Gamma_\mu = G_F^2 \hat{\Gamma}_{\mu,SM} + \Gamma_{\mu,ext} \ ,
\cr\cr
&&
\label{gamma_sme}
\eeqs
so 
\beqs
\frac{G_{F,app}^2}{G_F^2} &=& 1 + \frac{\Gamma_{\mu,ext}}{\Gamma_{\mu,SM}} =
                              1 + \frac{\Gamma_{\mu,ext}}{\Gamma_\mu - \Gamma_{\mu,ext}} \cr\cr
    &=& 1 + \frac{BR_{\mu,ext}}{1-BR_{\mu,ext}} \ . 
\label{gfsme}
\eeqs
Assuming that the BSM physics responsible for the additional contribution,
$\Gamma_{\mu,ext}$, to $\mu$ decay does not affect nuclear beta decays, then
the resultant apparent value of $|V^{'}_{ud}|^2$ obtained from the superallowed
nuclear beta decays would be given by $G_{F,app}^2 |V_{ud}'|^2 = G_F^2
|V_{ud}|^2$, i.e.,
\beq
|V^{'}_{ud}|^2=\frac{|V_{ud}|^2}{1+BR_{\mu,ext} }\ .
\eeq
For our present analysis, let us further assume that the BSM physics
leading to this value would not affect the decays used to determine
$|V_{us}|$ and $|V_{ub}|$. The apparent value of $\Sigma$, denoted
$\Sigma_{app}$, would then be
\beqs
\Sigma_{app} &=& |V_{ud}'|^2 + |V_{us}|^2 + |V_{ub}|^2 \cr\cr
&=& -BR_{\mu,ext}|V_{ud}'|^2 + |V_{ud}|^2 +  |V_{us}|^2 + |V_{ub}|^2 \cr\cr
&=& -BR_{\mu,ext}|V_{ud}'|^2 +  \Sigma \ .
\label{sigapp}
\eeqs
Assuming CKM unitarity, i.e., $\Sigma=1$, we then have
\beq
BR_{\mu,ext}=\frac{1-\Sigma_{app}}{|V_{ud}'|^2} \ .
\label{bmu}
\eeq
Presuming that this is responsible
for $\Sigma_{app}$ being less than unity and using the experimentally
determined value and uncertainty in Eq. (\ref{ckmsumsq}),
\beq
BR_{\mu,ext} < 1.3 \times 10^{-3} \ .
\eeq
%

% =================================================================

\section{Constraints from Leptonic $\tau$ Decays}
\label{tau_decay_section}

As with nuclear beta decay and the two-body leptonic decays of pseudoscalar
mesons, semihadronic $\tau$ decays have the simplifying property of only
involving a single leptonic charged-current vertex in their amplitudes, so one
may define an effective mass $m_{\tau,eff} = [\sum_i |U_{\tau,i}|^2
m_{\nu_i}^2]^{1/2}$. The best upper limit $m_{\nu_\tau,eff} < 18.2$ MeV (95\% C.L.)
\cite{Barate-aleph} comes from semihadronic $\tau$ decays.

As in the case of $\mu$ decay, one can analyze leptonic $\tau$ decays in the 
presence of possible sterile neutral emission; see Table II in \cite{shrock81b}
and also Ref. \cite{bryman_picciotto}. 
We denote the $\tau\to \nu_\tau e \bar \nu_e$ mode
as $\tau \to e$ and the  $\tau\to \nu_\tau \mu \bar \nu_\mu$ as $\tau \to
\mu$ for short and define a reduced, dimensionless decay rate $\bar\Gamma_{\tau
  \to \ell}$ via the equation
\beq
\Gamma_{\tau \to \ell } = \frac{G_F^2 m_\tau^5}{192\pi^3} \, 
(1+\delta_\alpha) \, \bar\Gamma_{\tau \to \ell}  \ 
\label{gammabar_tau}
\eeq
where we have used the fact that the leading-order correction, $\delta_\alpha$,
is mass-independent. In the Standard Model with neutrinos of negligible masses,
\beq
\Gamma_{\tau \to \ell,SM} = f(a_\ell^{(\tau)},0,0) \ . 
\label{gamma_tau_to_ell_sm}
\eeq
Using $a_e^{(\tau)} = 0.827 \times 10^{-7}$ and
$a_\mu^{(\tau)}=3.536 \times 10^{-3}$ in Eq. (\ref{fsm}), one has
$f(a_e^{(\tau)},0,0)=1-(0.662 \times 10^{-6})$ and
$f(a_\mu^{(\tau)},0,0)=0.9726$. 

With massive (anti)neutrino emission, we calculate
\begin{widetext}
  \beqs
\bar\Gamma_{\tau \to \ell} & = & 
(1-|U_{\ell 4}|^2)(1-|U_{\tau 4}|^2)f(a_\ell^{(\tau)},0,0) +
 (1-|U_{\ell 4}|^2)|U_{\tau 4}|^2 f(a_\ell^{(\tau)},0,a_{\nu_4}^{(\tau)})
    \cr\cr
 &+& |U_{\ell 4}|^2(1-|U_{\tau 4}|^2) f(a_\ell^{(\tau)},a_{\nu_4}^{(\tau)},0) +
            |U_{\ell 4}|^2|U_{\tau 4}|^2
            f(a_\ell^{(\tau)},a_{\nu_4}^{(\tau)},a_{\nu_4}^{(\tau)}) \ . 
  \label{gamma_tau_ell}
  \eeqs
\end{widetext}
Just as in Eq. (\ref{gamma_mu}) for $\mu$, the term involving emission of
$\nu_4 \bar\nu_4$ is negligibly small relative to the other terms because 
it involves the product of two small leptonic mixing matrix elements squared,
$|U_{\ell 4}|^2|U_{\tau 4}|^2$, and because of the greater kinematic 
suppression of the decay into $\nu_4 \bar\nu_4$ for substantial $m_{\nu_4}$;
one can therefore drop the
final term in Eq. (\ref{gamma_tau_ell}). The kinematic function $f(x,y,0)$ was
calculated in \cite{shrock81b}. It is worthwhile to inquire what can be learned
from a purely leptonic observable which can be calculated and measured to high
precision, namely
\beq
\frac{BR_{\tau \to e}}{BR_{\tau \to \mu}} \equiv R^{(\tau)}_{e/\mu}
\label{taubb}
\eeq
and the resultant ratio
\beq
\bar R^{(\tau)}_{e/\mu} = \frac{R^{(\tau)}_{e/\mu}}
                                 {R^{(\tau)}_{e/\mu,SM}} \ .
\label{rr_tau_emu}
\eeq
We comment below on studies that also include semihadronic $\tau$ decays.

Measurements of the individual branching ratios for 
$\tau\to \nu_\tau e \bar \nu_e$ and $\tau \to
\nu_\tau \mu \bar\nu_\mu$ have been carried out, with the results 
\cite{pdg}
\beq
BR_{\tau \to \nu_\tau e \bar\nu_e} = 0.1782 \pm 0.0004 
\label{br_tau_e}
\eeq
and
\beq
BR_{\tau \to \nu_\tau \mu \bar\nu_\mu} = 0.1739 \pm 0.0004 \ . 
\label{br_tau_mu}
\eeq
Experiments have also reported measurements of the 
ratio $R^{(\tau)}_{e/\mu}$; a global fit to the data yields the result
\cite{pdg,taunote_pdg} 
\beq
R^{(\tau)}_{e/\mu} = 1.024 \pm 0.003 \ . 
\label{taubb_exp}
\eeq
This is consistent with the theoretical SM prediction 
\beqs
&& R^{(\tau)}_{e/\mu,SM} \equiv \bigg (\frac{BR_{\tau \to e}}
                        {BR_{\tau \to \mu}} \bigg )_{SM} 
= \frac{f(a_e^{(\tau)},0,0)}{f(a_\mu^{(\tau)},0,0)} = 1.028 \ . \cr\cr
&& 
\label{taubb_sm}
\eeqs
The uncertainty in the theoretical prediction (\ref{taubb_sm}) is small
compared with the uncertainty in the experimental measurement
(\ref{taubb_exp}).  Note that the leading-order radiative correction term
$(1+\delta_\alpha)$ divides out in the ratio (\ref{taubb_sm}) since it is
mass-independent. Thus,
\beq
\bar R^{(\tau)}_{e/\mu} = 0.996 \pm 0.003 \ . 
\label{rr_tau_emu_exp}
\eeq
The simplest situation applies if $m_{\nu_4}$ is sufficiently large
that all of the decays $\tau \to \nu_4 e \bar\nu_j$ and 
$\tau \to \nu_4 \mu \bar\nu_j$, where $1 \le j \le 4$, are kinematically 
forbidden. In this case, 
\beqs
R^{(\tau)}_{e/\mu} &=& \frac{(1-|U_{e 4}|^2)f(a_e^{(\tau)},0,0)}
                       {(1-|U_{\mu 4}|^2) f(a_\mu^{(\tau)},0,0)}  \cr\cr
&=& \frac{(1-|U_{e 4}|^2)}{(1-|U_{\mu 4}|^2)} \, R^{(\tau)}_{e/\mu,SM} 
\quad\quad ({\rm no \ emission \ of} \ \nu_4) \ , \cr\cr
&&
\label{rem1}
\eeqs
i.e., $\bar R^{(\tau)}_{e/\mu} = (1-|U_{e4}|^2)/(1-|U_{\mu 4}|^2)$. 
Requiring that $\bar R^{(\tau)}_{e/\mu}$ not deviate excessively from 1
yields an upper bound on the magnitude of the difference $|U_{e 4}|^2
- |U_{\mu 4}|^2$, although this does not by itself provide separate
upper bounds on $|U_{e4}|^2$ or $|U_{\mu 4}|^2$.

Let us investigate a hierarchical lepton mixing situation in which
$|U_{\ell 4}|^2 << |U_{\tau 4}|^2$ for $\ell=e, \ \mu$. This is
effectively equivalent to using the upper limits $m_{\nu_e,eff} < 2$
eV and $m_{\nu_\mu,eff} < 0.19$ MeV \cite{pdg} to infer that these
have a negligible effect on the ratio $R^{(\tau)}_{e/\mu}$. Then
\begin{widetext}
\beq
R^{(\tau)}_{e/\mu} = \frac{(1-|U_{\tau 4}|^2)f(a_e^{(\tau)},0,0) + 
|U_{\tau 4}|^2 f(a_e^{(\tau)},a_{\nu_4}^{(\tau)},0)}
{(1-|U_{\tau 4}|^2)f(a_\mu^{(\tau)},0,0) + 
|U_{\tau 4}|^2 f(a_\mu^{(\tau)},0,a_{\nu_4}^{(\tau)})} 
= R^{(\tau)}_{e/\mu,SM} \Bigg [
\frac{(1-|U_{\tau 4}|^2) + |U_{\tau 4}|^2 
\bar f(a_e^{(\tau)},a_{\nu_4}^{(\tau)},0)}
     {(1-|U_{\tau 4}|^2) + |U_{\tau 4}|^2
\bar f(a_\mu^{(\tau)},a_{\nu_4}^{(\tau)},0)} \Bigg ] \ , 
\label{taubb_nu4}
\eeq
where we have used the symmetry (\ref{fsym}). Solving 
Eq. (\ref{taubb_nu4}) for $|U_{\tau 4}|^2$, we get
\beq
|U_{\tau 4}|^2 = \frac{ \bar R^{(\tau)}_{e/\mu}-1}
  {\bar R^{(\tau)}_{e/\mu}[1-\bar f(a_\mu^{(\tau)},a_{\nu_4}^{(\tau)},0)] - 
 [1-\bar f(a_e^{(\tau)}, a_{\nu_4}^{(\tau)},0)]} \ . 
\label{usqsol}
\eeq
\end{widetext}
With (\ref{rr_tau_emu_exp}), we obtain a 95 \% CL upper bound on $|U_{\tau
  4}|^2$ that extends down to below $10^{-2}$ as $m_{\nu_4}$ increases to 
1 GeV. More stringent constraints have been obtained from semihadronic decays
\cite{helo2011,abada,kobach_dobbs2015,gouvea}.

One can also use the measured branching ratios 
(\ref{br_tau_e}) and (\ref{br_tau_mu}) and the $\tau$ lifetime 
$\tau_\tau = (2.903 \pm 0.005) \times 10^{-13}$ s \cite{pdg} in
comparison with the decay rates calculated using the MuLan value for $G_F$ to
obtain limits on $m_{\nu_\tau,eff}$.  
The definition of $m_{\nu_\tau,eff}$ is more
complicated here than in nuclear beta decays and two-body leptonic decays of 
pseudoscalar mesons, where only a single charged-current vertex is involved, 
so $m_{\nu_e,eff} = [\sum_j |U_{ei}|^2 m_{\nu_j}^2]^{1/2}$ and 
$m_{\nu_\mu,eff} = [\sum_j |U_{\mu i}|^2 m_{\nu_j}^2]^{1/2}$, where the
sums include all neutrino mass eigenstates that lead to the respective 
outgoing charged lepton with an energy or momentum such that it is included in
the cuts used by a given experiment in its event reconstruction.  In contrast, 
for leptonic $\tau$ decays, the amplitudes involve two
charged-current vertices and hence products of lepton mixing matrices. If one
assumes that the $\nu_4$ is emitted via the $\tau-\nu_{\tau}$ charged-current
coupling, then only the $U_{\tau j}$ lepton mixing matrix element is relevant
in the amplitude, and one can express $m_{\nu_\tau,eff}$ in an analogous 
manner, as $m_{\nu_\tau,eff} = [\sum_j |U_{\tau i}|^2 m_{\nu_j}^2]^{1/2}$. 
Then, using the formulation in \cite{schael}, one finds calculated
values for the branching ratios (denoted by superscript $(c)$) of 
$R^{(c)}_{\tau \to e} = 0.17781 \pm 0.0031$ and 
$BR^{(c)}_{\tau \to \mu}=0.17293 \pm 0.00030$. Then, the ratios of 
experimental to calculated branching ratios are 
\beq
S_{\tau\to e}=   \Gamma_{\tau\to e}/\Gamma_{\tau\to e,SM}=1.022\pm 0.0028
\eeq
and 
\beq
S_{\tau\to \mu}=\Gamma_{\tau\to \mu}/\Gamma_{\tau\to \mu,SM}=1.0056 \pm 0.0029.
\eeq
Since the measured values exceed the calculated ones, we find 
the following 95\% C.L. interval for the physical regions for 
massive neutrino emission i.e. $S_{\tau\to e}<1$ and $S_{\tau\to \mu}<1$:
\beq
0.9964 < S_{\tau\to e} < 1
\label{Staue}
\eeq
and
\beq
0.9982 < S_{\tau\to\mu} < 1 \ .
\label{Staumu}
\eeq
Eqs. (\ref{Staue}) and (\ref{Staumu}) correspond to limits $m_{\nu_\tau,eff} <
38 $ MeV from $\tau \to \nu_\tau e \bar\nu_e$ and $m_{\nu_\tau,eff} < 26.8$ MeV
from $\tau \to \nu_\tau \mu \bar\nu_\mu$. These limits may be compared with the
current limit, $m_{\nu_\tau,eff} < 18.2$ MeV \cite{Barate-aleph}.

% ======================================================================

\section{Remarks on Some Other Particle and Nuclear Physics Constraints}
\label{other_constraints_section}

Sterile neutrinos with masses in the range considered here are subject to a
number of other constraints.  We begin with a remark on $K_{\ell 3}$ decays as
potential sources of constraints on sterile neutrinos. These decays include
$K^+ \to \pi^0 \ell^+ \nu_\ell$, $K^0_L \to \pi^+ \ell^- \bar\nu_\ell$, and
$K^0_L \to \pi^- \ell^+ \nu_\ell$, where $\ell=e, \ \mu$. Since these $K_{e3}$
decays are not helicity-suppressed, in contrast to $M_{e2}$ decays, where
$M=\pi^+, \ K^+$, etc., there is no associated enhancement of $K_{\ell 3}$
decays into a massive $\nu_4$ resulting from removal of helicity suppression,
as is the case in $M_{e2}$ decays.  These $K_{\ell 3}$ decays into a massive
(anti)neutrino are subject to the usual three-body phase space suppression. The
maximum $\nu_4$ masses in the $K_{e3}$, $(K^0_L)_{e3}$, $K_{\mu 3}$, and
$(K^0_L)_{\mu 3}$ decays are 358, 362, 253, and 252 MeV, respectively. This
mass range is already covered by the limits from peak search and branching
ratio constraints from $\pi_{\ell 2}$ and $K_{\ell 2}$ experiments.
Furthermore, the calculations of the rates for $K_{\ell 3}$ and
$(K^0_L)_{\ell 3}$ decays involve more uncertainty than for $\pi_{\ell 2}$
and $K_{\ell 2}$ because the hadronic amplitudes contain form factors whose
dependence on $q^2$ (where $q^\lambda= p^\lambda - p_\pi^\lambda$ is the
four-momentum imparted to the outgoing $\ell^- \nu_i$ or $\bar\ell^+ \nu_i$
pair) cannot be calculated from first principles.  (For a recent discussion of
parametrizations of these form factors, see \cite{pdg}.)  The resultant
uncertainty is only partially cancelled in ratios such as $BR((K^0_L)_{e
  3})/BR((K^0_L)_{\mu 3})$, since the $(K^0_L)_{e3}$ and $(K^0_L)_{\mu 3}$
involve different momenta transfers to the outgoing lepton pairs.

Next, it may be recalled that quite restrictive upper limits on mixings of
mainly sterile heavy neutrinos have also been obtained from time-of-flight
searches \cite{timingtest,gallas} and for neutrino decays
\cite{pdg,shrock81a,gr84,charm,ps191}. A recent search of this type is
\cite{t2k}. In the mass range of a few MeV, experiments have been performed to
search for the decay $\bar\nu_4 \to e^+ e^- \nu_e$ using $\bar\nu_e$ beams from
nuclear reactors \cite{oberauer87,kopeikin90,hagner95}. These eventually
obtained upper limits on $|U_{e4}|^2$ of $0.5 \times 10^{-2}$ at $m_{\nu_4}=1$
MeV down to $3 \times 10^{-4}$ for $m_{\nu_4}=4$ MeV, and then increasing to
$0.6 \times 10^{-2}$ for $m_{\nu_4}=9.5$ MeV \cite{hagner95}.  From
observations of the solar neutrino flux, the Borexino experiment has set upper
bounds $|U_{e4}|^2$ of $10^{-3}$ to $0.4 \times 10^{-5}$ for $m_{\nu_4}$ from
1.5 MeV to 14 MeV \cite{borexino1}.  However, since the conditions for the
diagonality of the neutral weak current are violated in the presence of sterile
neutrinos \cite{leeshrock77,sv80}, a sterile neutrino may decay invisibly, as
$\nu_4 \to \nu_j \bar\nu_\ell \nu_\ell$.  Other invisible neutrino decay modes
occur in models in which neutrinos couple to a light scalar or pseudoscalar
(for recent discussions and limits, see, e.g.,
\cite{kamlandzen_majoron}-\cite{archidiacono_hannestad2014} and references
therein). Consequently, because of their model-dependence, we do not use limits
on lepton mixing from neutrino decays here.

One can also check that a $\nu_4$ mass in the range
considered here would make a negligible contribution to decays such as $\mu^+
\to e^+ \gamma$. The branching ratio for $\mu^+ \to e^+ \gamma$ is 
\cite{p77}
\beq
BR(\mu^+ \to e^+ \gamma) = \frac{3\alpha_{em}}{32\pi}\, 
\Big (\frac{m_{\nu_4}}{m_W} \Big )^4 \, |U_{\mu 4}U_{e 4}|^2 \ . 
\label{br_mueg}
\eeq
For $m_{\nu_4}=100$ MeV, the factor multiplying $|U_{\mu 4}U_{e 4}|^2$
is $0.52 \times 10^{-15}$.  Given the experimental upper limit
$BR(\mu^+ \to e^+ \gamma) < 4.2 \times 10^{-13 }$ \cite{meg}, it is
clear that this does not yield a useful constraint on $|U_{\mu 4}U_{e
  4}|^2$.

A massive Dirac neutrino has a magnetic moment \cite{fs80} (see also
\cite{rs82}; recent reviews include \cite{gs_emrev,bkrev}).  Limits on such a
magnetic moment are commonly quoted for the interaction eigenstates, although
they are really properties of the mass eigenstates.  For one of the three SM
neutrino mass eigenstates $\nu_i$ in an active neutrino interaction eigenstate
$\nu_\ell$, where $\ell=e, \ \mu, \ \tau$, this is
\beqs
\mu_{\nu_i} &=& \frac{3eG_F m_{\nu_i}}{8 \pi^2 \sqrt{2}} \, \sum_\ell |U_{\ell i}|^2\cr\cr
&=& (3.2 \times 10^{-19}) \, \bigg ( \frac{m_{\nu_i}}{1 \ {\rm eV}} \bigg ) \,
\bigg [ \sum_\ell |U_{\ell i}|^2 \ \bigg ] \, \mu_B \ , \cr\cr
&&
\label{munu}
\eeqs
where $\mu_B = e/(2m_e)$ is the Bohr magneton. For a heavy (mostly sterile)
neutrino mass eigenstate with a mass in the range considered here, the
expression for $\mu_{\nu_42}$ is given by Eq. (\ref{munu}) with $i=4$. 
Thus, a Dirac $\nu_4$ with a mass of 5 MeV would have $\mu_{\nu_4} =
(1.6 \times 10^{-12}) \, [\sum_{\ell} |U_{\ell 4}|^2] \, \mu_B$ \cite{fs80}.

The upper limits conventionally quoted for the neutrino interaction eigenstates
are of order $10^{-10}- 10^{-11})\mu_B$ from reactor and accelerator
experiments, $3 \times 10^{-11}$ from the Borexino experiment \cite{borexino2},
and of order $(10^{-11}-10^{-12})\mu_B$ from limits on stellar cooling rates
\cite{pdg}. Since $|U_{e1}|^2$, $|U_{\mu 2}|^2$, and $|U_{\tau 3}|^2$ are
$O(1)$, there is not a large difference between the usually quoted upper limits
on ``$\mu_{\nu_e}$'', ``$\mu_{\nu_\mu}$'', ``$\mu_{\nu_\tau}$'', and the upper
limits on $\mu_{\nu_i}$ with $i=1, \ 2, \ 3$.

The situation is different for a heavy $\nu_4$ in the mass range considered
here.  When considering how these limits might apply to the $\nu_4$, however,
one must take into account the fact that there would be strong kinematic and
mixing-angle suppression or exclusion of the initial emission of the heavy
$\bar\nu_4$ in the beta decays that yield the $\bar\nu_e$ flux from a reactor,
and a $\nu_4$ with an MeV-scale mass would be kinematically forbidden from
being emitted in the $pp \to {\rm D} + e^+ + \nu_e$ reaction and the
electron-capture transition $e+{}^7{\rm Be} \to {}^7{\rm Li} + \nu_e$ in the
sun, since these have maximum energy releases of only 0.42 MeV and 0.86 MeV,
respectively.  Hence, one cannot necessarily apply the constraints on neutrino
magnetic moments from reactor antineutrino and solar neutrino experiments to a
heavy neutrino. Similarly, the constraint from stellar cooling is not directly
relevant here because it only applies to neutrino mass eigenstates $\nu_j$ with
masses $\lsim 5$ keV so that a plasmon in the star would be kinematically able
to produce the $\bar\nu_j \nu_j$ pair \cite{pdg}.

Finally, we comment on how a heavy neutrino could affect Higgs decays.  Ref.
\cite{invisible} pointed out that the Higgs boson could have decays to
invisible final states, and calculated rates for several of these, including
decays to neutrinos.  Currently, all of the decay branching ratios of the Higgs
are in agreement with SM predictions, but these allow for a substantial
branching ratio into invisible modes, $BR(H \to {\rm invisible}) \lsim 20$ \%
\cite{pdg,atlas_inv,cms_inv}.  The way in which the diagonal and nondiagonal
couplings of neutrinos to the Higgs boson are related to the couplings $U_{\ell
  4}$ that enters in the weak charged current depends, in general, on details
of a given model.

% =====================================================================

\section{Conclusions}
\label{conclusion_section}

One of the most important outstanding questions in nuclear and particle physics
at present concerns whether light sterile neutrinos exist.  In this paper we
have presented a detailed analysis yielding new upper bounds on the squared
lepton mixing matrix elements $|U_{e4}|^2$ and $|U_{\mu 4}|^2$ involved in the
possible emission of a mostly sterile neutrino mass eigenstate, $\nu_4$, from
analyses of a number of nuclear and particle decays. A brief report on the
upper bounds on $|U_{e4}|^2$ was given in \cite{ul}. We have used recent
advances in the precision of measured ${\cal F}t$ values for a set of
superallowed nuclear beta decays to improve the upper limits on $|U_{e4}|^2$
obtained from these beta decays for a $\nu_4$ with a mass in the range of a few
MeV. From analyses of the ratios of branching ratios
$R^{(\pi)}_{e/\mu}=BR(\pi^+ \to e^+ \nu_e)/BR(\pi^+ \to \mu^+ \nu_\mu)$,
$R^{(K)}_{e/\mu}$, $R^{(D_s)}_{e/\tau}$, $R^{(D_s)}_{\mu/\tau}$, and
$R^{(D)}_{e/\tau}$, and from $B_{e 2}$ and $B_{\mu 2}$ decays, we have derived
upper limits on couplings $|U_{e4}|^2$ and $|U_{\mu 4}|^2$. Our bounds on
$|U_{e4}|^2$ cover most of the $\nu_4$ mass range from approximately 1 MeV to 1
GeV, and in several parts of this range they are the best bounds for a Dirac
neutrino that do not make use of model-dependent assumptions on visible
neutrino decays.  We have also obtained a new upper bound on $|U_{\mu 4}|^2$
from a $\pi_{\mu 2}$ peak search experiment searching for $\nu_4$ emission via
lepton mixing and have updated existing upper bounds on $|U_{\mu 4}|^2$ in the
MeV to GeV mass range.  New experiments to search for $D_s^+ \to e^+ \nu_4$ and
$D^+ \to e^+ \nu_4$ are suggested. These, as well as a continued search for
$B^+ \to e^+ \nu_4$ and $B^+ \to \mu^+ \nu_4$ decays, would be valuable and
could further improve the bounds.  In addition, we examined limits on
$|U_{e4}|^2$ obtained from examining pion beta decay and showed that they are
less stringent than those from superallowed beta decay in the same $\nu_4$ mass
range. As part of the analysis, we updated constraints from CKM unitarity on
sterile neutrinos.  In addition, we examined correlated constraints on lepton
mixing matrix coefficients $|U_{e4}|^2$, $|U_{\mu 4}|^2$ and $|U_{\tau 4}|^2$
from analyses of leptonic decays of heavy-quark pseudoscalar mesons, from $\mu$
decay, and from leptonic $\tau$ decays.

% ======================================================================

\begin{acknowledgments}

  We thank J. Hardy, W. Marciano, and M. Ramsey-Musolf for discussions on the
  extraction of $|V_{ud}|$ from superallowed nuclear beta decays. We thank
  Changzheng Yuan and Guang Zhao for discussions about BES III data analyses
  and Jose Benitez and Vera Luth for information about BABAR data analyses.
  This work was supported by the Natural Sciences and Engineering Research
  Council and TRIUMF through a contribution from the National Research Council
  of Canada (D.B.)  and by the U.S. NSF Grants NSF-PHY-1620628 and 
  NSF-PHY-1915093 (R.S.).

\end{acknowledgments} 

% =====================================================================

% ========================================================================


\begin{thebibliography}{99}
\bibitem{cosm}
%
There is an extensive literature on these cosmological constraints and on 
models that are designed to evade one or more of them. Some of the many papers
are listed in \cite{cosm1}-\cite{cosm5}.
\bibitem{ml2radcor}
%
  The dominant term in the radiative correction in $R^{(M)}_{\ell/\ell'}$ is
  $\delta_{RC}=3(\alpha_{em}/\pi)\ln(m_{\ell}/m_{\ell'})$ \cite{earlyrad,ms93}.
  (The radiative correction in the limit where $m_{\nu_4}$ approaches its upper
  kinematic limit is irrelevant, since the overall rate for the decay vanishes
  in this limit.) To evaluate the term $\delta_{RC}$ that enters in the
  expression (\ref{rmsm}) for $BR(M^+ \to \ell^+ \nu_\ell)/BR(M^+ \to \ell'^+
  \nu_{\ell'})$ in the case of heavy-quark pseudoscalar mesons $M$, we use the
  results of Ref. \cite{ms93}, replacing $m_\pi$ with $m_M$, where $M=D, \ D_s,
  \ B$ and replacing $m_\rho$ with $m_{M^*}$, where $M^*=D^*, \ D_s^*, \ B^*$.
\bibitem{neutralinos}
Subsequent to \cite{megg,kpigg}, lower limits on neutralinos
in possible supersymmetric extensions of the SM have increased
substantially, so that decays such as these would be
kinematically forbidden.
\bibitem{taunote_pdg}
%
The Particle Data Group lists two results for 
$BR(\tau \to \nu_\tau \mu \bar\nu_\mu)/BR(\tau \to \nu_\tau e \bar\nu_e)$, 
namely a fit, equal to $0.9762 \pm 0.0028$ and an average, equal to 
$0.979 \pm 0.004$ \cite{pdg}.  We have used the fit value, since it is more
precise.
\end{thebibliography}
\end{document}